# Space-Time Diffusion of Ground

# and Its Fractal Nature

*V. Shiltsev*

*Fermi National Accelerator Laboratory, PO Box 500, Batavia, IL 60510, USA*

*Abstract*

We present evidences of the diffusive motion of the ground and tunnels and show that if systematic movements are excluded then the remaining uncorrelated component of the motion obeys a characteristic fractal law with the displacement variance $dY^2$ scaling with time- and spatial intervals $T$ and $L$ as $dY^2 \propto T^\alpha L^\gamma$ with both exponents close to 1 ($\alpha \approx \gamma \approx 1$). We briefly describe experimental methods of the mesa- and microscopic ground motion detection used in the measurements at the physics research facilities sensitive to the motion, particularly, large high energy elementary particle accelerators. A simple mathematical model of the fractal motion demonstrating the observed scaling law is also presented and discussed.

# 1 Introduction

Motion of the ground was always of practical interest because of the scare of earthquake-induced damage and concerns about structural stability of buildings due large movements. In recent decades, development of large-scale facilities for scientific research also confronted the issue of very tight tolerances on element's positions in the presence of microscopic motion of the ground. The most notable examples are gravitational wave detectors [1-3] and high energy particle accelerators [4-7]. In the gravitational wave detectors, the ground vibrations transferred to the motion of the mirrors in the arms of interferometers are one of the sources of noise limiting minimum detectable strain. In the accelerators, motion of numerous focusing magnets disturbs trajectories of tiny charged particle beams and, thus, affecting machine performance. Given the tight tolerances on positioning, quite sophisticated measurement, stabilization and correction/alignment systems are routinely employed there [8]. To design such systems one relies on certain phenomenological models of the ground motion which should predict the expected displacement of the ground $Y(t,s)$ depends on the time interval $t$ and distance between the points of control $s$. The spatial scales of interest $L$ for these physics instruments range from several meters to dozens of km and the time intervals of interest $T$ range from ms to years.

The instruments for the microscopic ground motion measurements have been originally developed for geophysics research, currently many of them are made easily applicable for other purposes and commercialized. Among widely used at the large physics facilities are optical interferometers, stretched wires and hydrostatic level systems (HLS) [9], laser position trackers [10], and geophones [11]. They are quite capable to detect the movements over the above noted scales of $L$ and $T$ even under very quiet conditions.

Ambient ground motion has three distinct components – periodic motion (e.g. due to Earth tides, seasonal changes, etc), systematic drifts or trends (e.g due to temperature or air pressure variations, precipitation history, etc) and stochastic movements [12]. The stochastic component usually is less correlated in space, less persistent in time and less predictable than the first two while not necessarily smaller in amplitude, thus, often posing the biggest concern. Space-, time- or space-time variograms can be used to describe average characteristics of the motion $Y(t,s)$ :

$$
\begin{aligned}
&< dY^2(t,L) >=< \left( Y(t,s+L) - Y(t,s) \right)^2 > \\
&< dY^2(T,L) >=< \left( Y(t+T,s+L) - Y(t+T,s) - Y(t+T,s) + Y(t,s) \right)^2 >
\end{aligned}
\tag{1}
$$

where the brackets <…> denote averaging over continuous or discrete time series and $T$ and $L$ are lags in time and space. Below we present and discuss evidences that the stochastic component of the ground motion can be described as *diffusion in both time and space* and has a characteristic fractal law variogram :

$$< dY^2(T,L) > \propto T^\alpha L^\gamma \qquad (2)$$

with both exponents close to 1 ($\alpha \approx \gamma \approx 1$) over wide ranges of time- and space-intervals. Corresponding power spectral density (PSD) $P(\omega,k)$ in frequency $\omega = 2\pi f$ and spatial wave-number $k = 2\pi/\lambda$ for such a process scales as:

$$P(\omega,k) \propto \frac{1}{\omega^\beta k^\delta} \qquad (3)$$

with exponents $\beta = \alpha + 1$ *and* $\delta = \gamma + 1$ (detail discussion on mathematical methods of geophysical time series analysis can be found in [12]).

Power-law scaling of separate temporal or spatial variograms of the ground motion, i.e., dependencies of the type $<dY^2(T, L=const)> \propto T^\alpha$ and $<dY^2(t=const, L)> \propto L^\gamma$ have been long known to geophysicists, see, e.g. [13], [14], but it was high precision studies of dynamics numerous measurement points for large accelerators where simultaneous space- and-time diffusion was observed for the first time. An empirical *ATL law* [15] was proposed to summarize the experimental data, according to which the rms relative

displacement $dY$ of the points separated by a distance $L$ grows with the time $T$ as:

$$< dY^2 > = A\,TL \qquad (4)$$

where $A$ is a site dependent constant of the order of $10^{-5\pm1}$ $\mu m^2/(s{\cdot}m)$. Such a wandering of the ground elements takes place in all directions. As long as the diffusive coefficient $A$ is small the diffusion presents only a tiny contribution to the ground motion. For example, in the time period of 1 hour the amplitude of the absolute surface motion (e.g. measured by seismometer) could be as big as 100 $\mu m$, while the *ATL* estimates relative displacement of about 1 $\mu m$ for the points 30 m apart. One would not worry about this contribution except it describes very important, at least for accelerators, *uncorrelated* background on top of the larger amplitude ground movements correlated in time and space. The later includes, but not limited to, low frequency seismic waves, tides, an ambient low-frequency ground motion generated by local sources such as wind, air pressure variation, temperature gradients, ground water, precipitation, etc. Obviously, the ATL law is a particular case of the more general equation (1). The PSDs of the ATL-type motion in the frequency and the wave-number domains scale as:

$$P(f) = \frac{AL}{2\pi^2 f^2}, f > 0 \quad P(k) = \frac{2AT}{k^2}, k > 0 \ (5).$$

This article reviews the evidences of the space-time diffusion of the ground surface or tunnel. In Section 2, we discuss the measurements made at the particle accelerators with use of standard alignment instrumentation, describe briefly the impact of misalignments on the beams in accelerators and present evidences of the beam orbit diffusion caused by diffusion of elements' positions. Section 3 contains review results of various geophysical studies made either at the accelerator facilities, or at the sites of future accelerators, or at the geophysics labs. We summarize all the measurements and discuss the limits of validity of the space-time ground diffusion laws in Section 4 and present a simple numerical model of the fractal ground motion which generated the landscape evolution according to the empirical law.

## 2 Ground and Beam Orbit Diffusion in Accelerators

### 2.1 Impact of Ground Motion on Operation of Accelerators

For the purposes of this study, particle accelerators can be considered as sequence of linear focusing elements (magnetic lenses) arranged either in a circle (circular accelerators) or in a line (linear accelerators). In an ideal accelerator with perfectly aligned magnetic elements, the beam orbit passes through the centers of the lenses magnets. Any alignment error results in the beam orbit distortion. If the distortions are large compared to apertures of the lenses or the size of the vacuum chambers or the size of a linear focusing field areas, then they become an obstacle for successful operation of the machine and must be corrected – either with use of electromagnetic orbit correctors or by means of mechanical realignment which brings the centers of the focusing lenses back to their ideal positions [16]. In large accelerators, such as 6.3-km circumference proton-antiproton Tevatron Collider (Fermilab, Batavia, IL, USA), 27-km circumference proton-proton Large Hadron Collider (LHC at CERN, Switzerland), 6.3 km circumference proton-electron collider HERA at DESY (Hamburg, Germany), and 25-50 km long future electron-positron Linear Colliders, which have many hundreds of magnetic elements, the motion of the ground and corresponding displacements of the magnets are the most important source of the beam orbit distortions. It has to be noted that the biggest effect is produced by uncorrelated relative motion of the neighboring focusing elements while very long-wavelength movements are practically unimportant, and, for example, accelerators are not sensitive to their global displacements as a whole [6], [7]. Orbit distortions from numerous uncorrelated sources add in quadrature and, thus, the rms distortion of the beam orbit due to the ATL-law type ground motion (4) in a circular accelerator with circumference $C$ can be approximated as [17]:

$$< dY_{orbit}^{2} > \approx \kappa A T C \qquad (6),$$

that shows that larger orbit drifts are expected at larger accelerators. The numerical factor in (6) $\kappa \approx 2\text{-}5$ depends on the design of the beam focusing optics. Typically, the ground motion effects start to be of a serious concern for accelerators at the amplitudes of the uncorrelated motion from a fraction of a micron to a dozen of microns, depending on the accelerator parameters and types. For accelerators which collide tiny size beams the final focusing magnet stability tolerances could be as tight as microns to few nanometers [7]. Because of the concerns with the magnet position stability, large accelerators are usually been installed inside deep concrete-and-steel enforced tunnels (typical diameters/sizes of the order of 5-8 m at the depth from 10 to 100 meters) at the location with known good and stable geology.

## 2.2 Orbit Drifts in Large Accelerators

To a greater or lesser extent long-term orbit drifts are seen at all accelerators and machine operators or/and automatic correction systems counteract the drifts. As large colliding beam facilities are particularly sensitive to the orbit motion, some extended investigations of the issue have been carried out there. In this section we present observations of the beam orbit drifts in several large accelerators – HERA (Germany), TRISTAN (Japan), Tevatron (US) and LEP (Switzerland). Detailed parameters of these machines can be found in corresponding references below.

### 2.2.1 Orbit Drifts in HERA Proton-Electron Collider

HERA is a high energy accelerator in Hamburg (Germany), which was in operation as proton-electron collider in 1992-2007. The circumference of HERA is 6.3 km. The facility is located in an underground tunnel in a depth of approximately 25 meters below the surface. It consisted of two independent accelerators-storage rings for 30 GeV electrons and 820 GeV (since 1998 - 920 GeV) protons installed in the same tunnel (the height difference between electron and proton beam is 0.8 m, focusing optics lattice are very different).

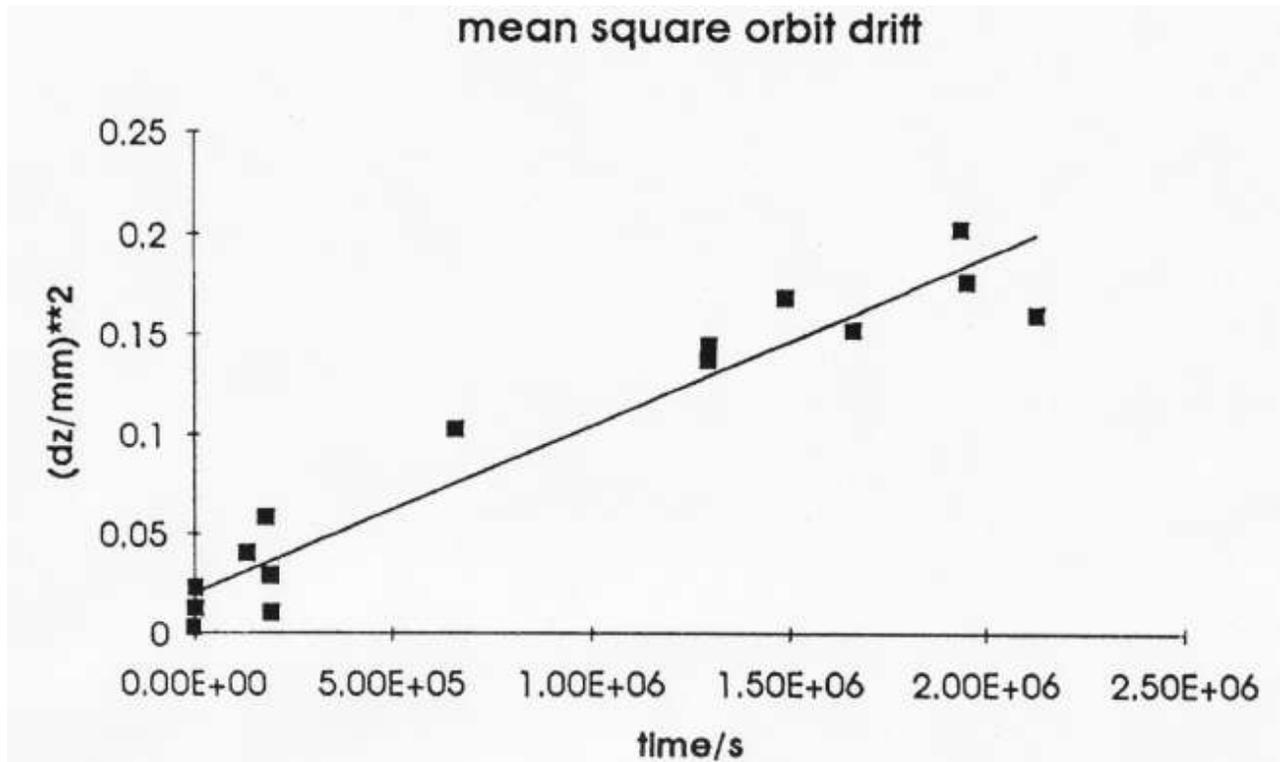

Fig.1: Mean square difference of vertical orbit distortions in the HERA electron ring vs time interval duration obtained from data stored during 1993 operation data [18].

Fig. 1 from [18] shows the mean square of the HERA electron ring vertical orbit drifts accumulated after various time intervals (up to 1 month) and detected by 288 beam position monitors located about 23 m from each other all over the circumference. One can see that the variance of the distortions grows approximately linearly in time $<dY_{orbit}{}^2(T)> = a+bT$ with $a=0.02$ mm$^2$ and $b=8 \cdot 10^{-8}$ mm$^2$/s. Here, the constant $a$ accounts for the noise of measurements, while the slope $b$ gives an estimate of the diffusive ground motion constant $A_{HERAe} \approx 4 \cdot 10^{-6}$ µm$^2$/s/m if one uses the optics coefficient $\kappa \approx 3.1$ for the HERA-e in Eq.(5).

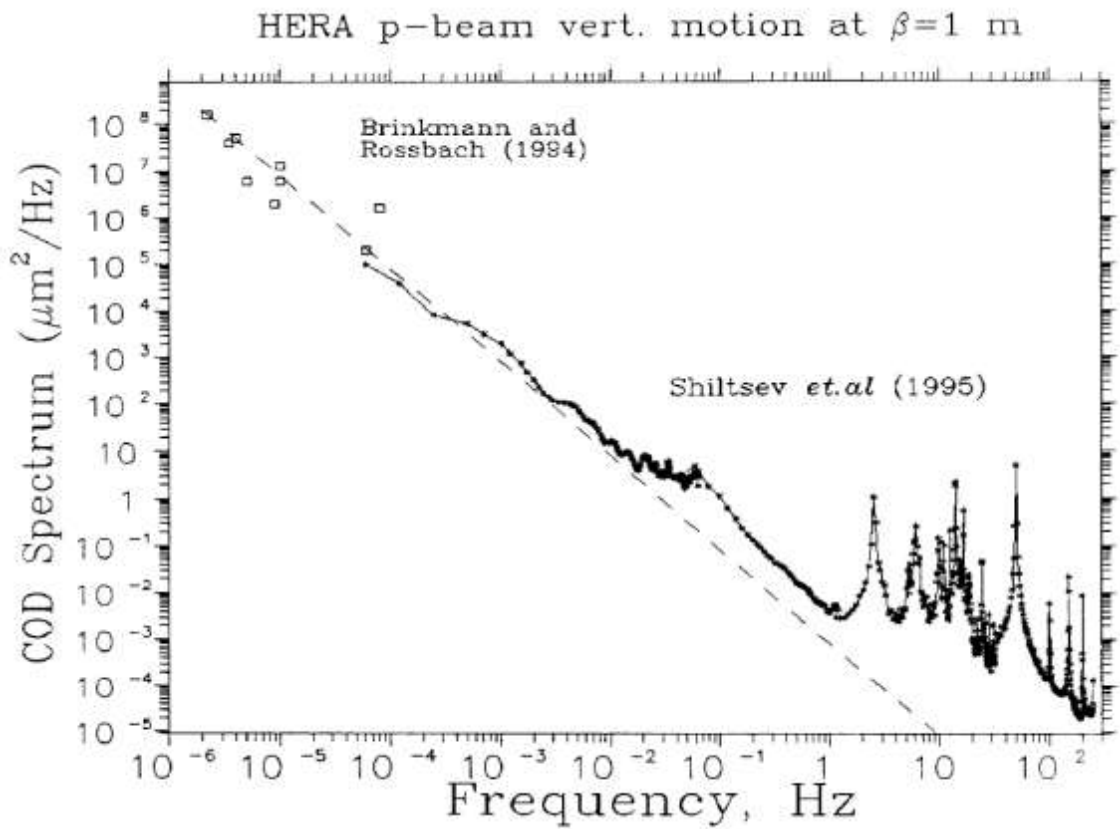

Fig.2:  PSD of the HERA proton orbit vertical motion normalized to a specific location of the ring. Dashed line is for the *ATL* expectation [18, 17].

Analysis of the vertical motion in the other (proton) ring is summarized in the Power Spectral Density (PSD) shown in Fig.2. The squares at lower frequencies represent the Fourier spectra of the proton orbit differences from different running periods of the accelerator [18]. The procedure was to measure the closed orbit position at all 131 BPMs in the HERA-proton machine and subtract the result from a previous one to obtain  the difference orbit, indicating any eventual orbit drift. The analysis of difference orbits was limited to time intervals of about 5 days maximum during which no intentional change of the closed orbit occurred. Continuous line represents the Fourier spectrum of readings from one specific beam position monitor in the accelerator [17].   As continuous observations were performed repetitively within several hours of the proton beam lifetime, the lowest frequency of this spectrum is about 0.5 mHz. Series of peaks in the spectrum above 1 Hz are due to cultural seismic noise which is quite prominent in a big city like Hamburg. The dashed line in Fig.2 shows the PSD scaling $P_{orbit}(f) >= 8 \cdot 10^{-4}$ $[\mu m^2 s]/f^2$ as expected from the *ATL* law with the constant   $A_{HERAp} \approx 8 \cdot 10^{-6}$ $\mu m^2/s/m$  which fits very well the data in the range of frequencies from $2 \cdot 10^{-6}$

Hz to about $2 \cdot 10^{-2}$ Hz. In time domain such a PSD corresponds to irregular noisy "random walk"-like proton orbit drifts over the time intervals few some minutes to several days. The PSD power-law fit results in the exponent of $\beta$=1.95±0.2. Mechanical motion of the focusing magnets was found to be the reason of the HERA orbit drifts, as other sources - long term drifts of orbit corrector strengths and low-frequency noises of the BPMs- were negligible.

### 2.2.2 Orbit Drifts in TRISTAN and KEK-B Positron-Electron Colliders

TRISTAN is a high energy accelerator in Tsukuba (Japan), which was in operation as positron-electron collider in 1986-1998. Its tunnel has about 3.0 km circumference, has 0.8 m thick concrete walls and set at a depth of approximately 12 meters below the surface. The energies of the beams of positrons and electrons were up to 32 GeV. Long term 8 GeV beam orbit drifts over several periods of a few days each have been reported in Ref.[19] and are shown in Fig.3. Full circles in the figure are the rms values of the beam positions $x_i$ in all $N=392$ BPMs while the open circles represent the rms of the position *changes* during operation cycles between successive corrections of the orbit, i.e. $\sigma=(\Sigma(x_i-x_{i0})^2/N)^{1/2}$.

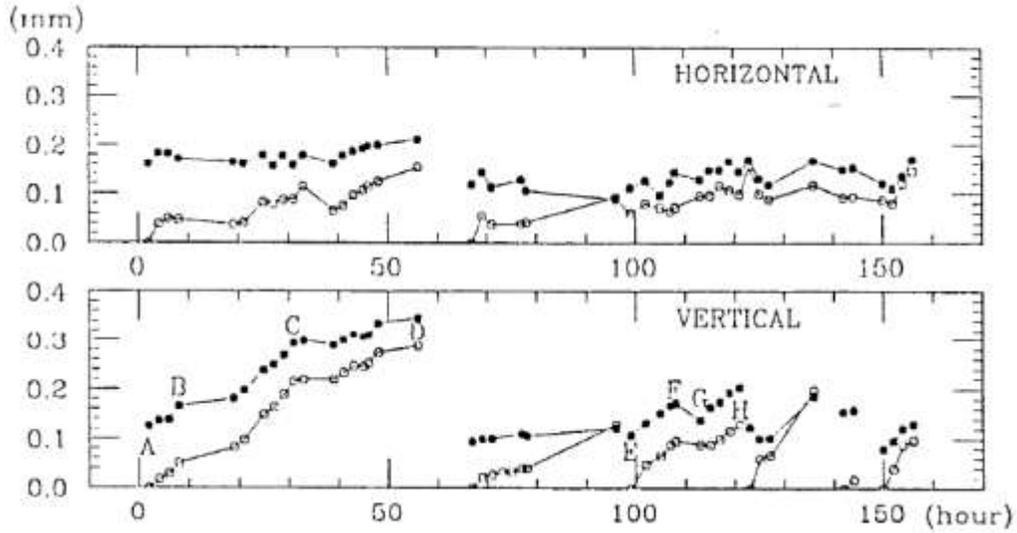

Fig.3: Changes of rms vertical and horizontal orbits in TRISTAN ring (from Ref. [19]).

Note that the horizontal COD is smaller than the vertical one. At large orbit distortions, the beam current circulating in the accelerator degraded significantly so that a correction of the orbit was needed toward the "ideal" orbit (sharp drops at points D, E, H, and some others in Figure 3).

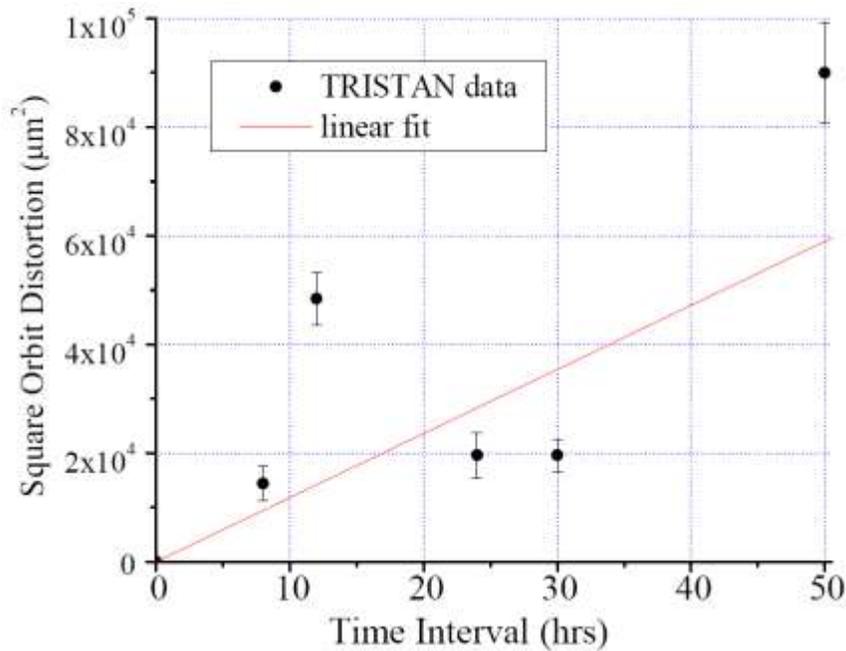

Fig.4: Variance of the TRISTAN orbit variations [17].

Analysis of the data presented in Figure 3, shows that the variance of the COD grows with the time [17] – see Fig.4 – and can be approximated by a linear fit (6) with coefficient $A_{TRISTAN}=(27\pm7)\cdot10^{-6}$ μm$^2$/s/m .

After the end of the TRISTAN operation, a new higher performance KEK-B positron-electron collider was built in the same tunnel and started its opera  tion  in 1999. That collider consists of two intersecting rings set side-by-side – one for 8 GeV electrons and another for 3.5 GeV positrons. Tight sub-mm control of  the ring's 3-km circumference is critical for the collider operation. Fig. 5 below shows 4 month record of the positron ring circumference change [20].

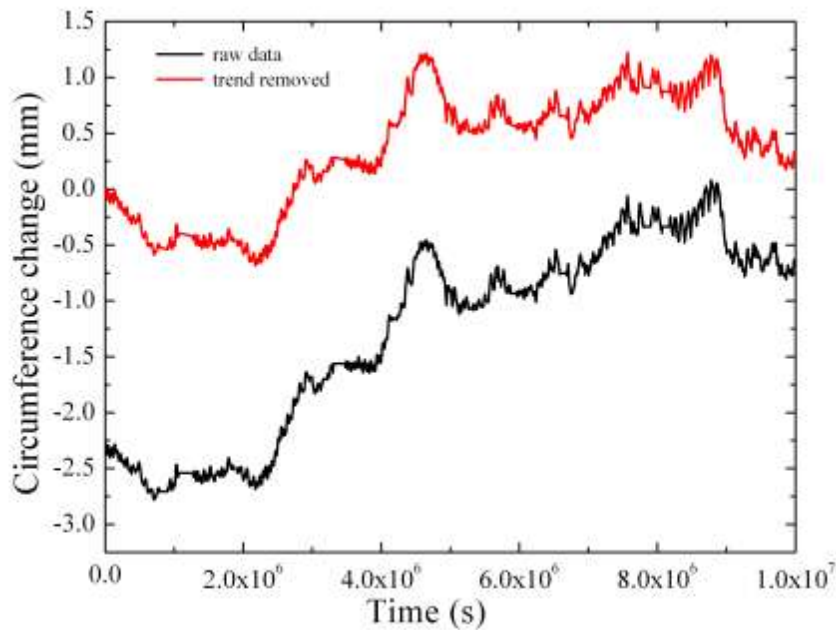

Fig.5:  KEK-B circumference variations from March 1 to June 30, 2002 [20].

If  linear trend is excluded from the data (see upper and lower curves in Fig.5) then the variogram  (1) of the circumference change $\Delta C$ after a time interval $T$ scales linearly with $T$ - see Fig.6 - as expected from the ATL law $<\Delta C^2>=2ATC$ with $A_{KEK}=(27\pm3)\cdot10^{-6}$ μm²/s/m – in a remarkable agreement with the TRISTAN orbit drift analysis results presented above.

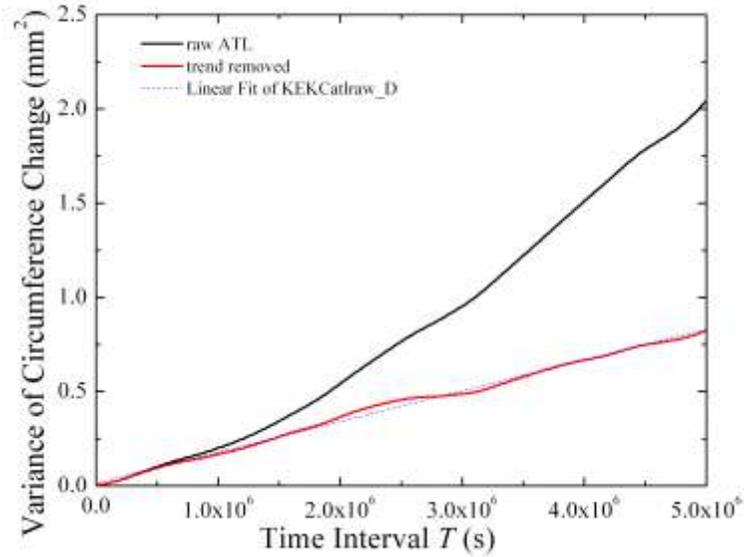

Fig.6:  Variance  of the KEK-B circumference variations; black line is for raw data, red line is for the data with linear trend subtracted, dashed line is a linear fit.

PSD of the circumference change is  presented in Fig. 7 and shows distinctive peaks at frequencies of  ~2/day (some 15 μm changes due to ground expansion due to solar and  lunar  tides)  and  some  30  μm  peak  due  to  daily  temperature  changes.  The circumference  also  found  changing  due  to  air  pressure  variation,  especially  during the time when a typhoon hit the area (not in Fig.3).   At very low frequencies less than $10^{-5}$ Hz , the PSD scales approximately as $1/f^{2.1\pm0.2}$ , also in decent agreement with Eq.(5).

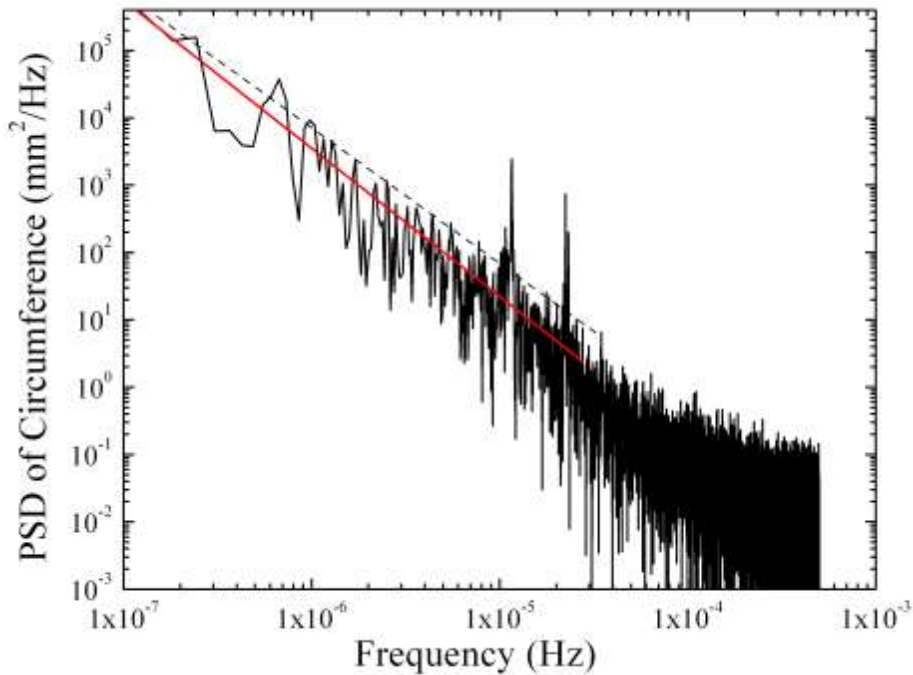

Fig.7:  Spectrum of the KEK-B circumference variations [20]. Dashed line is for the
ATL-law scaling $8.2/f^2$; solid red line is for a power law fit $(1.74 \pm 0.2)/f^{2.21 \pm 0.07}$.

### 2.2.3 Orbit Drifts in Tevatron Proton-Antiproton Collider

Tevatron Collider is currently (2009) the world's highest energy accelerator for high

energy physics research with beams of  980 GeV  protons and antiprotons circulating

in opposite directions  inside the same set of  774 bending magnets and 216 focusing

magnets. It is located in Batavia, IL (USA) in a 6.3 km circumference tunnel at

approximately 7 m below the surface. The motion of the tunnel floor translates into

motion of focusing magnets and further translates into movement of the beams. For

effective operation of the Collider, the beam orbit motion must be stabilized to within

0.1mm by means of the automatic orbit correction system.  Without such a system orbit daily changes can easily reach 0.2-0.3 mm as indicated in Fig.8 and as much as 0.5-1 mm over the periods of 2-4 weeks [21].

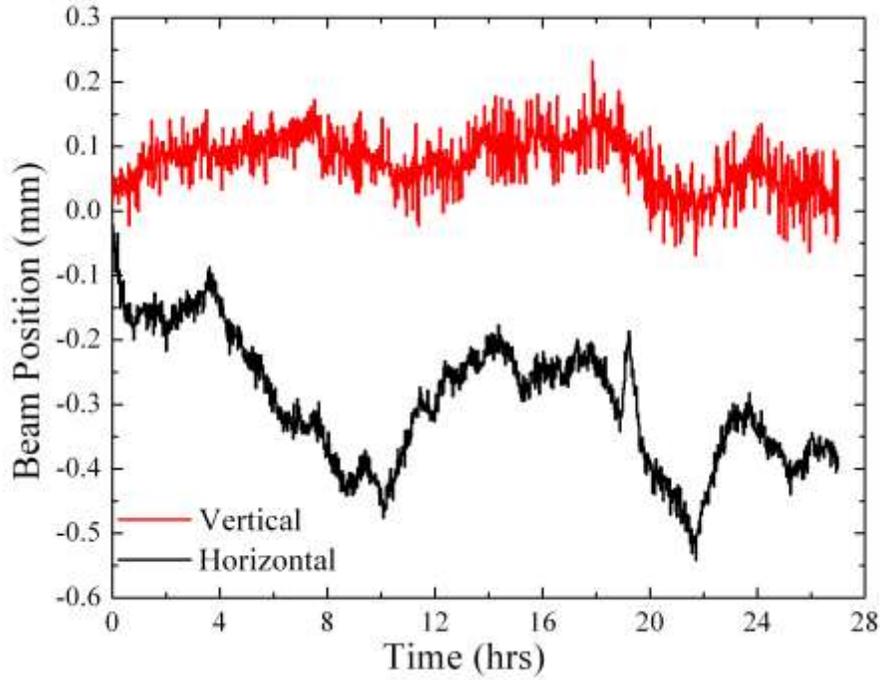

Fig.8: Horizontal and vertical orbit motion as measured by one of the beam position monitors in the Tevatron [21].

Besides the 12- and 24-hour variations associated with the tides and daily temperature effects, the orbit motion has a diffusive component. To separate it, one can compute the variance of *the second differences <ddY²(T)>* which is equal to :

$$< ddY^2(T) >=< \left( dY(t) - 2dY(t+T) + dY(t+2T) \right)^2 > \quad (7).$$

It is easy to see that contrary to variance of the (first) difference (1), effectively filters linear trends and slow periodic variations out. Indeed, for the process which contains a linear trend, a periodic component, a diffusive ATL-like component and truly uncorrelated noise (e.g. due to measurement errors) *dY(t)>=Et+Fsin(ωt)+(ATL-like diffision)+(noise with rms of G)* one gets :

$$< dY^2(T) >= E^2T^2 + 2F^2 \sin^2(\omega T/2) + ATL + 2G \quad (8a),$$

$$< ddY^2(T) >= 8F^2 \sin^4(\omega T/2) + 2ATL + 6G \quad (8b).$$

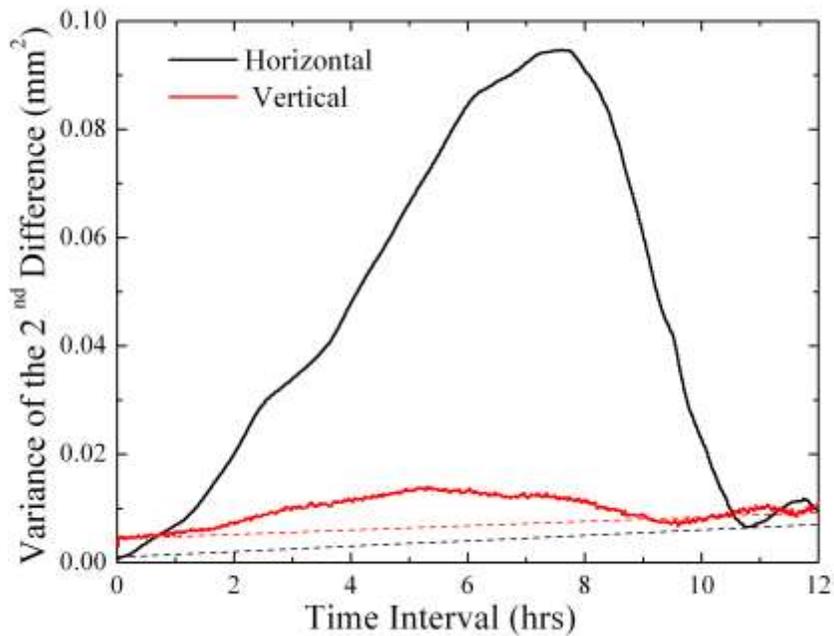

Fig.9: Tevatron Proton orbit 2nd difference variance. Dashed lines are linear fits of the ATL-like component of the variance.

The result of such analysis for the Tevatron orbit drift data is shown in Fig.10. One can see that both horizontal and vertical variances have significant diurnal (tide) components. The ATL-diffusion components scale linearly with the time lag $T$ and are indicated by dashed lines which have the of slopes of 0.0027±0.0003 mm$^2$ over 12 hours (horizontal) and .006±0.001 mm$^2$ over 12 hours (vertical). The diffusive coefficient A can be calculated from (8b) and (6) taking into account that beam optics factors $\kappa$ are different for horizontal and vertical planes [21], so $A_{Tevatron\ V} = (2.6±0.3)·10^{-6}$ $\mu m^2/s/m$ and $A_{Tevatron\ H} = (1.8±0.2)·10^{-6}$ $\mu m^2/s/m$.

*2.2.3 Orbit Drifts in CERN's Large Electron-Positron Collider (LEP) and Super-Proton Synchrotron (SPS)*

Large Electron-Positron collider (LEP) was the world's highest energy electron-positron collider under operation in European Organization for Nuclear Research (CERN) in Geneva, Switzerland in 1989-2000. Energy of the beams varied to as much as 104 GeV. 3368 bending magnets of LEP deflected the particles and kept them in orbit. There were also 816 focusing magnets and 700 orbit correctors. The 26.7 km circumference tunnel of LEP has eight straight sections and eight arcs and lies between 45 m and 170 m below the surface on a plane inclined at 1.4% sloping towards the Léman Lake. Approximately 90% of its length is in molasse rock, which has excellent characteristics for this application, and 10% is in limestone under the Jura mountain. Internal tunnel diameter: 3.8 m in the arcs. 4.4 and 5.5 m in the straight sections depending on the plant installed in them.

As for other accelerators we considered above, stability of the beam orbit was essential for successful operation of the collider. Motion of few very strong superconducting focusing magnets correlated with temperature variations at the magnet support structure was found to be main source of ~3 mm vertical beam orbit movements [22]. Employment of local orbit correctors allowed to reduce this effect by an order of magnitude. The residual orbit motion was found variance growing linearly with time interval – see Fig. 10. Applying the ATL law fit of Eq.(6) with coefficient $\kappa$ numerically evaluated in [23], one can estimate the diffusion constant $A_{LEP} = (10.9 \pm 6.8) \cdot 10^{-6}$ μm$^2$/s/m [24].

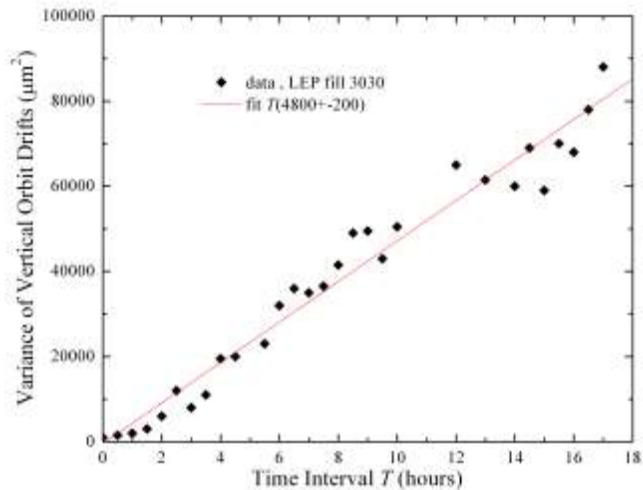

Fig.10: Variance of the LEP vertical orbit distortions vs time interval $T$ with effects of movements of the strongest focusing magnets removed (from Refs. [22, 24]).

Similar analysis has been extended for 30,000 orbits were recorded while LEP was colliding beams for its experiments in 1999 [23]. The orbit data was analyzed to reconstruct the orbit drifts that were compensated by the LEP slow orbit feedback and to remove the effects due to the earth tides, motion of few very strong superconducting focusing magnets mentioned above and other known intentional corrections implemented to optimize the accelerator operation.

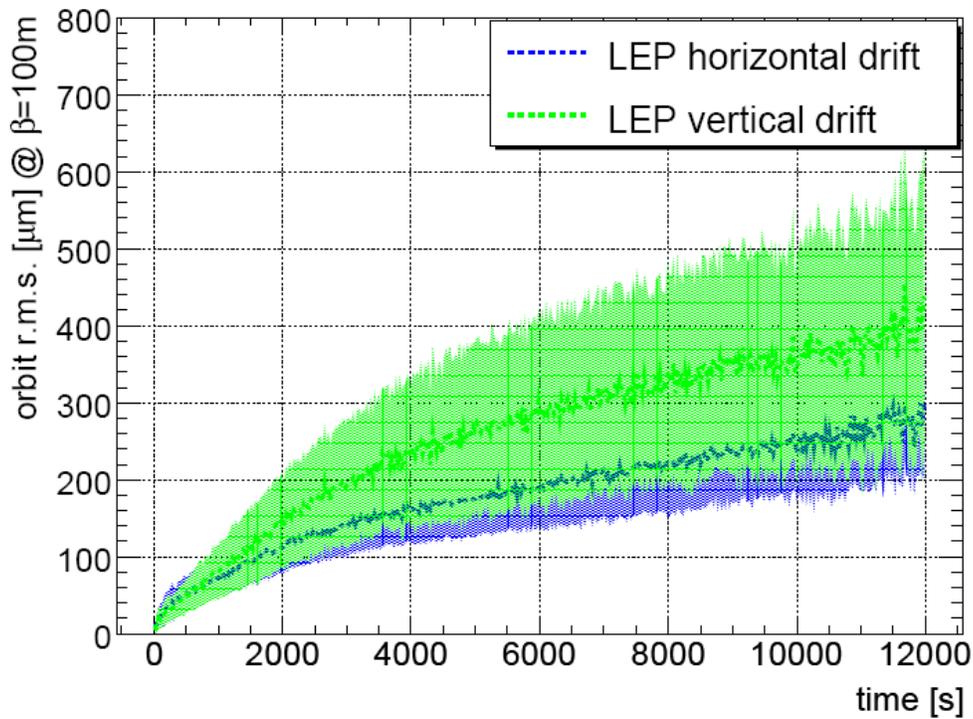

Fig.11: RMS vertical and horizontal LEP beam orbit drifts during 1999 operation. The $\sigma \propto T^{1/2}$ growth with time interval $T$ is visible (from Ref.[23]).

Figure 11 shows the orbit r.m.s. $\sigma_{V,H}$ normalised to an effective "average" monitor location in the ring. The data can be very well fitted by $\sigma_V = (3.6 \pm 1.5)[\mu m]\ T^{1/2}\ [s]$ and $\sigma_H = (2.56 \pm 0.7)$ [$\mu m$] $T^{1/2}$ [s] (note significant 30-40% spread in the data). Such a scaling is predicted from Eq.(6) and the diffusion coefficients can be calculated taking into account known coefficients $\kappa_{V,H}$ [23]. It is noted in Ref.[23] that since the influence of other (unknown) effects cannot be fully excluded, then following estimates should be considered only as upper limits for the diffusive ground motion constants $A^*_{LEPv} = (38 \pm 23) \cdot 10^{-6}$ $\mu m^2$/s/m and $A^*_{LEPh} = (32 \pm 19) \cdot 10^{-6}$ $\mu m^2$/s/m.

We believe that one of such effects which was not properly accounted in Ref.[23] is regular periodic orbit distortions due to the Earth tides. The above considered Tevatron orbit variations - Fig.9 – set an example which shows the tides, if not properly excluded from the data, can increase formally calculated diffusion coefficient by a factor of 2 to 10. It was reported in [25] that the tidal deformations of the Earth's crust do cause a 1 mm variation in the circumference of LEP. Variations of the orbit distortions over the time intervals of about 3 hours (considered in the Fig.11 data) can be as big as 10-30% of that, thus, possibly dominating the rms orbit analysis. In addition to the periodic tidal variations, slow systematic seasonal changes of the LEP circumference of 2 mm have been observed. These movements might also affect the orbit analysis. They are particularly pronounced after important rainfall and might be produced by an expansion of the earth or by a pressure due to underground water levels (sponge effect) [25].

Yet another accelerator at CERN, named Super Proton Synchrotron (SPS) has a circumference of about 6.9 km and an average depth of about 50m. Its tunnel (as well as the LEP one) is embedded in the *Molasse*, a soft tertiary sandstone on top of a hard rock basin found in the region. The *Molasse* mainly consists of clay and limestone eroded from the surrounding Jura and the Alps and is covered by the *Moraine*, a loose and permeable more recent quaternary erosion from the Jura. In 2004 long-term SPS orbit stability measurements were performed with beams of protons with energies up to 270 GeV. Figure 12 from Ref.[23] shows power spectra of the vertical beam motion of a 270GeV and 26GeV beams that was sampled by a monitor with about 2 μm r.m.s resolution (seen as white noise above 0.1Hz).

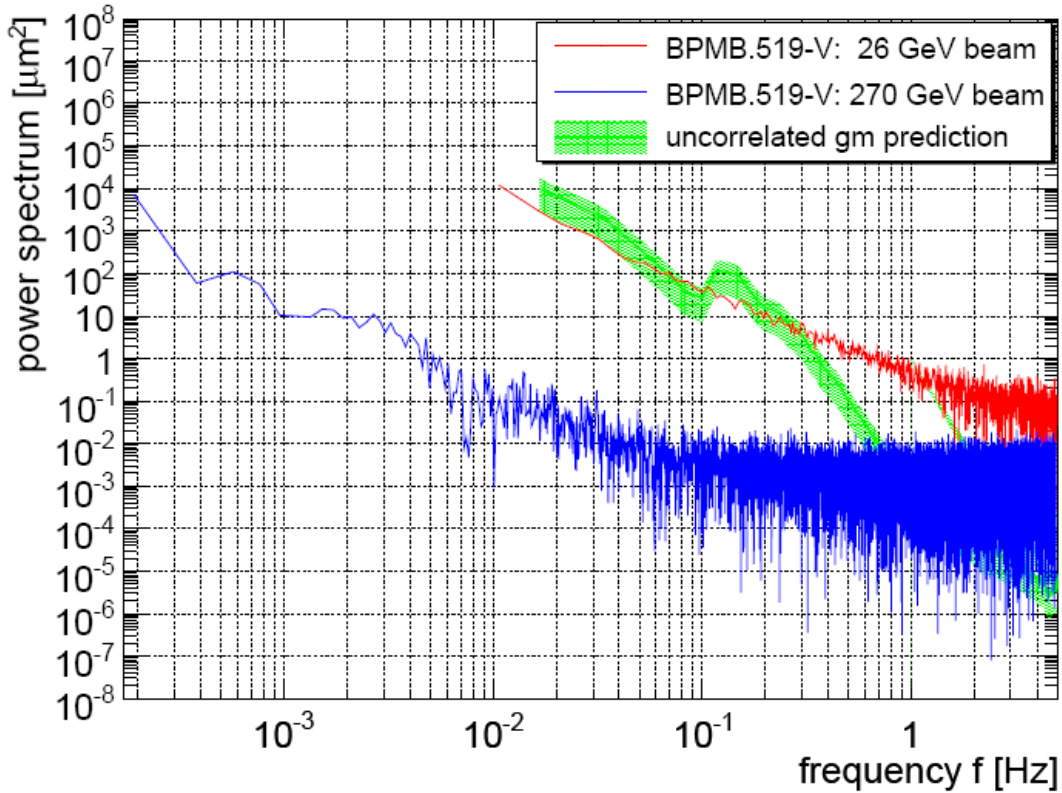

Fig.12: Power spectra of orbit movement at 26GeV and 270GeV in the SPS (from Ref.[23]).

The 26GeV data are thought to be dominated by slow drifts of the magnetic fields rather than by ground motion. The 270 GeV data shows characteristic ATL-law spectrum scaling of $1/f^2$. Using a pre-calculated vertical orbit sensitivity factor $\kappa$ for the SPS and fitting the observed orbit drifts spectra, the following SPS ground motion coefficient estimate can be obtained $A_{SPS} = (6.3 \pm 3) \cdot 10^{-6}$ μm$^2$/s/m.

### 2.3 Ground Diffusion in the Accelerators Alignment Data

Despite having sophisticated orbit correction systems, all accelerators undergo regular realignment of the magnets positions back their ideal values. That allows to reduce greatly the dependence on the correction systems and helps to maintain stable operation of the facilities over periods of many years. Modern commercial instruments, e.g. laser trackers, for geodetic survey and alignment allow to achieve accuracies of a fraction of a mm over distances of a km and their description can be found elsewhere (see, e.g., Ref. [8]). In this section we present analysis of long term ground motion drifts as observed during the realignment of large accelerators.

*2.3.1 Long-Term Motion of LEP Magnets*

Several times a year, positions of more than 700 focusing magnets of the LEP were measured and restored back to their prescribed values to follow an ideal smooth curve" . Results of the LEP magnets elevations measurements in 1993-1994 [26] are shown in Fig.13.

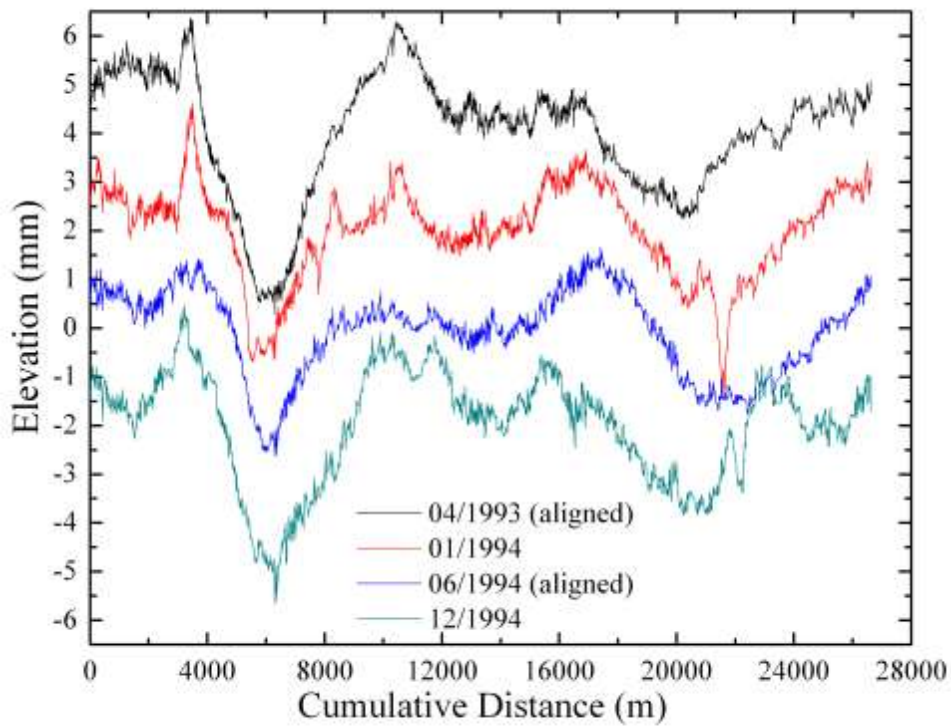

Fig.13. Elevations of the CERN's LEP focusing magnets measured in 1993-94 [26] vs cumulative distance along the ring (i.e. the point at 0 m placed close to the point at 26.7 km).

The average tilt of 1.4% was subtracted from the data. For the purpose of the presentation in one Figure, the four curves are vertically separated by 2 mm from each other. The top line in Fig.13 shows vertical positions the magnets in April 30, 1993, just after making the realignment of the accelerator to a smooth curve. The roughness of this curve is thought to be mostly due to the instrumentation accuracy. Some 9 months after the April 1993 realignment, on January 28, 1994, the positions had been re-measured – see the 2[nd] from the top line. One can see that the line is more rough and several peaks have appeared, the biggest are around 3500 m and 21500 m which are the regions of systematic long-term drifts due to well known geological instability. Then, the realignment had been done and the LEP magnets elevations as measured June 6, 1994 are presented in the 3[rd] from the top line. Major peaks are now smoothed. Six month after, in December 1994, they reappear, see the bottom line in Fig.13, together with other smaller changes. Further analysis and data processing made in Ref.[17], include: 1) 1 km pieces of the LEP circumference around 3500 m and 21500 m were excluded from the analysis; 2) as one is not interested in the smooth spatial curves, the lowest five Fourier harmonics were subtracted from the data. Now, the variances of the first difference $<dY^2(L)>=< (dY(l)-dY(l+L))^2>$ have been calculated as where brackets $< . . . >$ denote averaging over all possible pairs of the magnets

distanced by *L*. The results are presented in Fig.12 where the straight lines represent liner fits :

$$April\ 30, 1993 \quad <dY^2(L)>_I = (0.0147 \pm 0.0014) + L \cdot (1.63 \cdot 10^{-4} \pm 2.4 \cdot 10^{-6}) \quad (9a),$$

$$January\ 28, 1994 \quad <dY^2(L)>_{II} = (0.0218 \pm 0.005) + L \cdot (3.72 \cdot 10^{-4} \pm 8.9 \cdot 10^{-6}) \quad (9b),$$

$$June\ 6, 1994 \quad <dY^2(L)>_{III} = (0.0001 \pm 0.0043) + L \cdot (2.36 \cdot 10^{-4} \pm 7.3 \cdot 10^{-6}) \quad (9c),$$

$$December\ 1994 \quad <dY^2(L)>_{IV} = (0.017 \pm 0.005) + L \cdot (3.42 \cdot 10^{-4} \pm 9.2 \cdot 10^{-6}) \quad (9d).$$

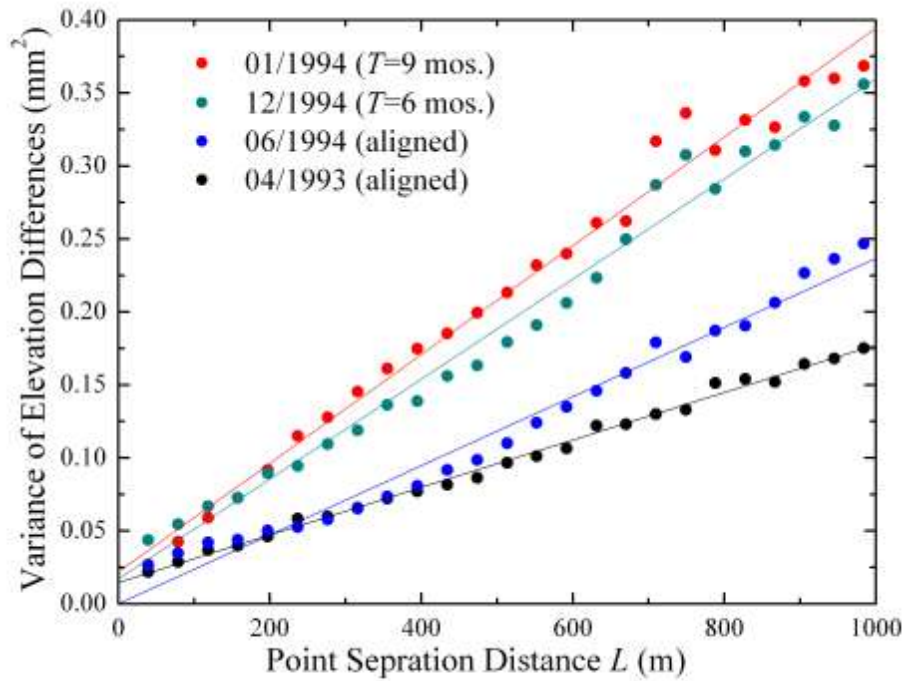

Fig.14. The variance of relative displacement of the CERN LEP magnets vs. the distance between them *L* (from Ref.[17]).

One can see that for $L < 1000$m, the variances for just-realigned accelerator $<dY^2(L)>_I$ and $<dY^2(L)>_{III}$ are 1.5-2 times less than what is measured after several months without alignment. It has to be noticed that the variance grows linearly with $L$ even right after the alignment. That is because of the method of survey when the alignment is made sequentially – one segment of the machine after another – and the random errors of the position measurement of a given magnet with respect to the previous one add up like a random walk. Such a random walk error can be estimated by the closure errors of about 2 mm over the entire circumference (measured at different periods) that is equivalent to 0.14mm$^2$/km – in a good agreement with the analysis shown in Fig.14. The increase of the variance after the time interval (the top two lines) over the instrumentation noise (the bottom tow lines) should be assigned to the ground diffusion. Again, assuming validity of the *ATL* law, one gets two estimates of the diffusion constant *A:*

$$A_{II-I} = \frac{<dY^2(L)>_{II} - <dY^2(L)>_I}{L \cdot 9 months} = (9.0 \pm 0.5) \cdot 10^{-6} \; \frac{\mu m^2}{s \cdot m} \quad (10a),$$

$$A_{IV-III} = \frac{<dY^2(L)>_{IV} - <dY^2(L)>_{III}}{L \cdot 6 months} = (6.8 \pm 0.8) \cdot 10^{-6} \; \frac{\mu m^2}{s \cdot m} \quad (10b).$$

which are remarkably close to each other. Therefore, the LEP alignment data demonstrate that the variance of the relative displacements in time scales proportionally to the distance between the points. Six-year elevation changes of the LEP magnets in 1993-1999 have been analyzed in Ref.[27]. It was shown that after exclusion of the linear trends and systematic drifts from the data, the remaining random diffusion can be described by the ATL law with coefficient $A_{LEP} = (2.9 \pm 0.6) \cdot 10^{-6}$ $\mu m^2/s/m$.

### 2.3.2 Motion of CERN's Super Proton Synchrotron Magnets

The noted above CERN's Super Proton Synchrotron (SPS) was constructed in mid-1970s and has 6.9 km circumference. There are 744 bending magnets and $N$=216 focusing magnets placed practically uniformly over the ring. Primary data from an optical survey shown in Fig.15 represent the vertical displacements of the magnets relative to the theoretical "ideal" position of 1976. These values were measured three times at about three years intervals: in 1985, 1988 and 1991 – with estimated accuracy of about few dozens of micrometers.

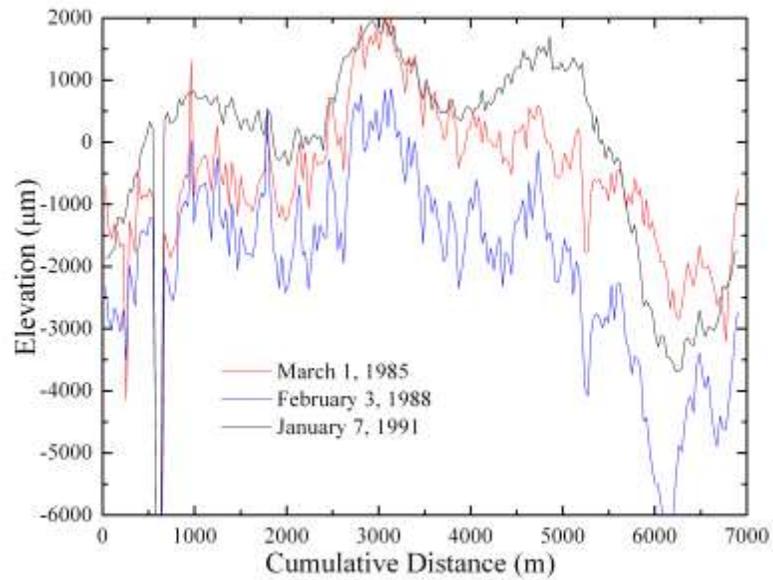

Fig.15. Displacements of the CERN's Super Proton Synchrotron magnets measured
in 1985, 1988 and 1991 along the circumference ring; the point at 0 m placed close
to the point at 6912 m (courtesy of J.-P.Quesnel of the CERN's Survey Group).

These data were processed the way similar to the one as for the LEP
alignment data discussed above, so, for example, the values for several
magnets around 600 m and few were not taken into considerations as these
magnets were intentionally displaced during the period.

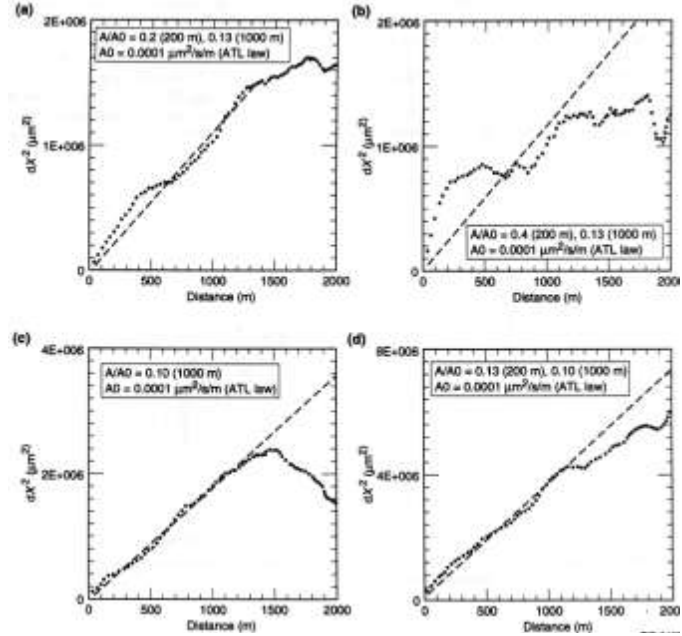

Fig.16. The variance of the relative vertical displacement of the SPS magnets after various time intervals vs. distance between the points of the position survey *L:*

a) - 3 years (1985~1988), b) - 3 years (1988-1991), c) - 6 years (1985-1991), d) - 12 years (1976-1988) (from Ref.[28]).

The variances of the relative vertical displacements of the magnets versus distance *L* are presented in Fig.16 from [28] together with linear fits (dashed lines) according to the ATL law with diffusion coefficients of $20 \cdot 10^{-6}$ $\mu m^2/s/m$, $40 \cdot 10^{-6}$ $\mu m^2/s/m$, $10 \cdot 10^{-6}$ $\mu m^2/s/m$ and $13 \cdot 10^{-6}$ $\mu m^2/s/m$ for time intervals of 1985-1988, 1988-1991, 1985-1991 and 1976-1988, correspondingly. It has to be emphasized that the time intervals vary from 3 years to 12 years, and nevertheless the diffusive constants are almost the same. An average value of the coefficient for the SPS data is thus $A_{SPS}$

$=(14\pm5)\cdot10^{-6}$ $\mu m^2/s/m$. Note, that a power-law fit $<dY^2(L)>\propto L^{\gamma}$ with exponent $\gamma$ less than 1 might better describe the variances than the linear fit.

### 2.3.3 Tevatron Alignment Data Analysis

Alignment system of the Tevatron Collider employs more than 200 geodetic "tie rods" installed in the concrete tunnel wall all over the ring , approximately 30 m apart.

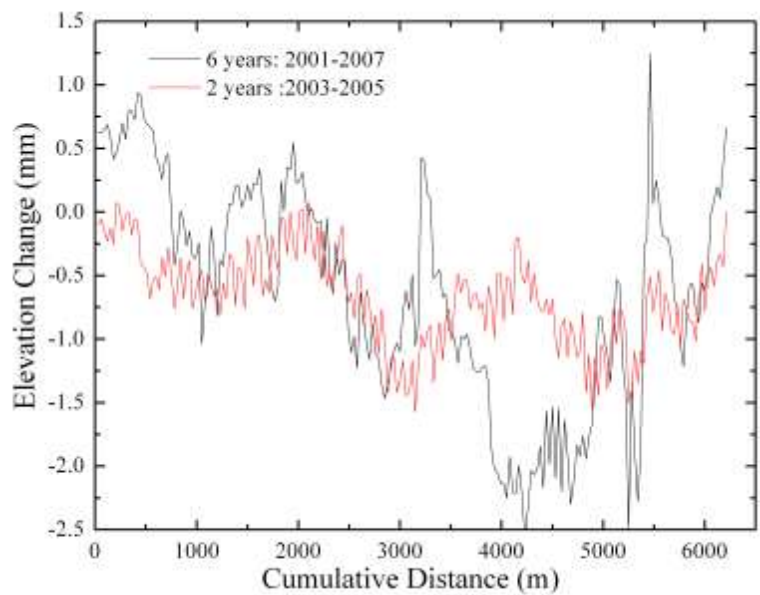

Fig.17: Vertical displacement of more than 200 "tie rods" in the Tevatron tunnel over the period of 2003-2005 and a 6 year period of 2001-2007 (data courtesy of J.Volk and Fermilab's Alignment Group).

Position of the magnets is regularly locally referenced with respect to the rods while positions of the rods are routinely globally monitored. The rods elevations data are available for the years of 2001,2003,2005,2006 and 2007. Fig.17 shows the change of the elevations around the ring accumulated over two intervals – 2 years (2003-2005) and 6 years (2001-2007). One can see that longer term motion has larger amplitude. The variance $<dY^2(L)>=<(dY(z)-dY(z+L))^2>$ of the displacements has been calculated and averaged over all possible time intervals. E.g. there are two 1-year intervals (that is 2005-2006, 2006-2007), three 2-year intervals (2001-2003, 2003-2005, 2005-2007), etc, and one for the 6-year interval 2001-2007. The results for 1-year changes and for the 6-year change are shown in Fig.18. A remarkable difference between the two plots is that 1 year variance scales linearly only up to $L \approx 900\ m$ and does not depend on $L$ beyond that scale, while the 6 years variance grows all the way to distances as large as 1800 m [29]. Such a behavior indicates independence of the displacements of the rods located more than 900 m apart on the time scale of a year, and existence of a significant level of interdependence of the motion of distanced rods at the times as long as 6 years. The calculated variances for all possible time difference can be well approximated by linear fits $<dY^2(L)> =a+bL$ over distances less than 900 m and the slopes (fit parameters $b$ with the error bars) are plotted in Fig.19.

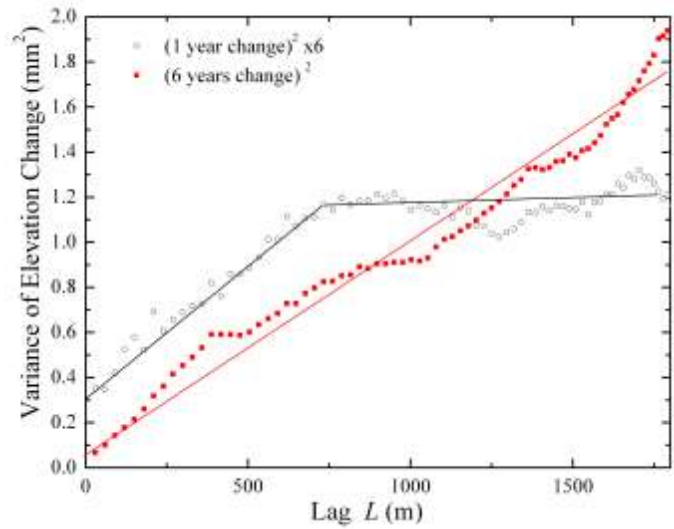

Fig.18: Variances of the averaged Tevatron tie rod vertical displacements over time intervals of 1 (multiplied by 6) and 6 years vs the distance $L$ (from Ref.[29]).

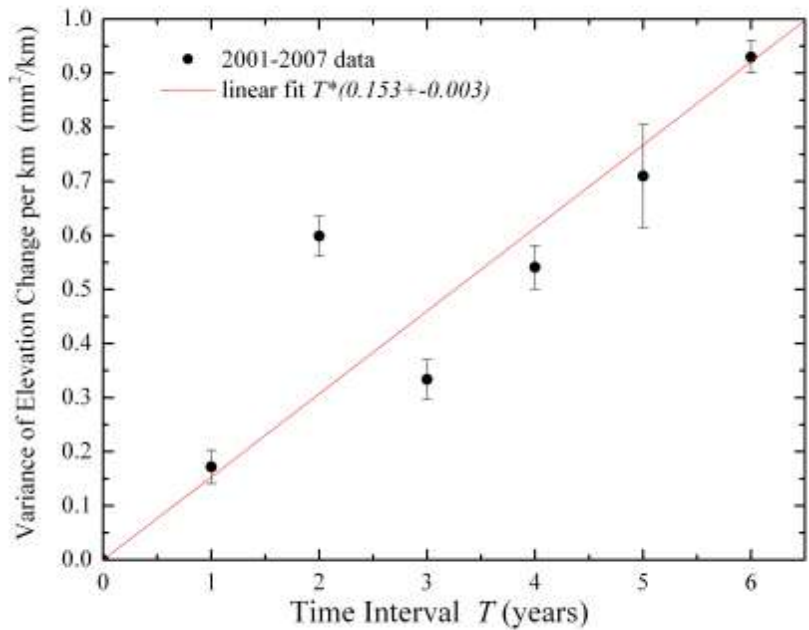

Fig.19: Variances of the Tevatron alignment rods displacements per unit distance vs the time interval between the measurements (see text, from Ref.[29]).

One can see that the variance per unit distance grows with the time interval between the measurements, and can be approximated by a linear fit $b(T) = cT$ with $c=0.153\pm0.004$ $[mm^2/km/year]$. Such dependence is in accordance with the ATL law with coefficient $A_{Tevatron} = c = (4.9\pm0.13)\cdot10^{-6}$ $\mu m^2/s/m$ [29].

*2.3.4 Alignment Data on Ground Motion in Other Accelerators*

The variance of the 1985-1988 SPS elevation changes are compared with the alignment data from several other accelerators sites in Fig.20. Because of the different times of observations for these data, they are presented as functions of the variance of displacement divided by the time of observations vs. distance $L$ between the points of the ground. For comparison, the *ATL* law scaling with coefficient $A = 100\cdot10^{-6}$ $\mu m^2/s/m$ is also shown by a dashed line. That line well approximates the theodolite measurements of vertical movements of few dozen surface monuments along a 2 km long straight line at the UNK collider construction site (Protvino, Moscow region, Russia) made over time interval $T$ of about 2 yr.

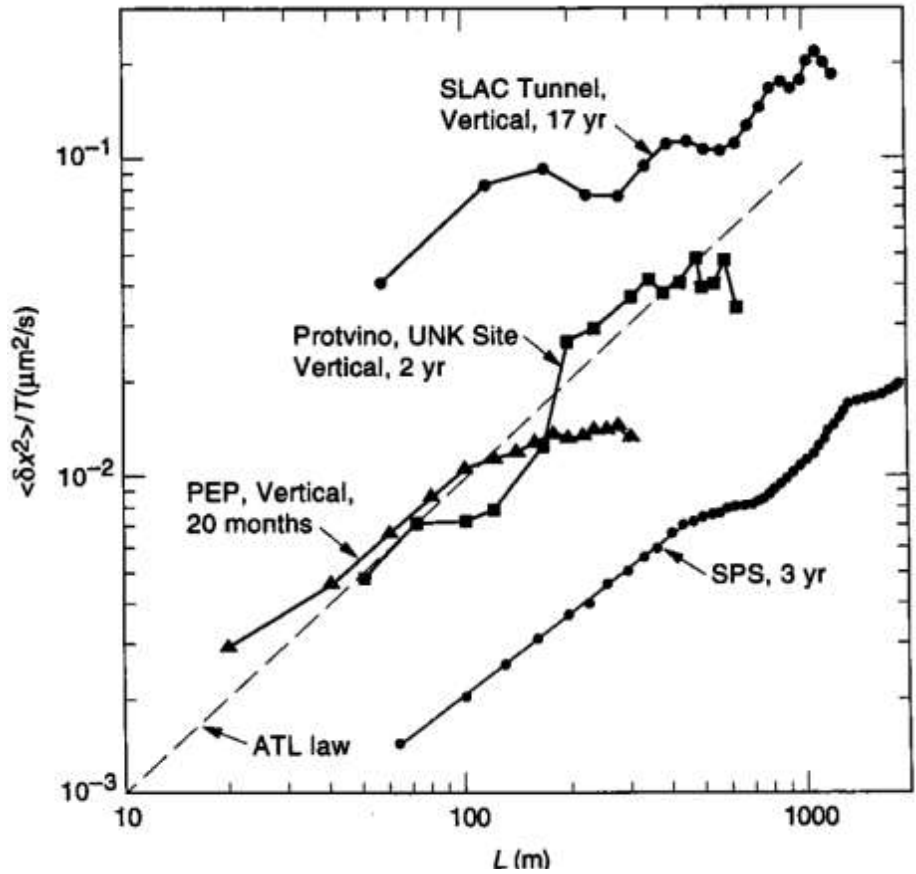

Fig.20: Variances of the accelerator magnet displacements per unit time vs distance

for the SLC, UNK, PEP and SLAC tunnel (see text, from Ref. [28]).

The other two lines represent the data of the measurements made at the

Stanford Linear Accelerator Center (SLAC) accelerators: one for the 2 km

circumference PEP accelerator magnet displacements during 20 months

(1989-1991) and another is for very long term displacements over 17 years of the elements in a 2-mile linear accelerator tunnel. These tunnels in SLAC sit on or are mined in grey unweathered well cemented tertiary myocene sandstone. Possibly due to "cut and cover" construction method and smaller depth, the SLAC linac tunnel demonstrates faster diffusion than the PEP tunnel - the coefficients are $A_{SLC} = (200\pm100)\cdot10^{-6}$ $\mu m^2/s/m$ and $A_{PEP} = (100\pm50)\cdot10^{-6}$ $\mu m^2/s/m$ correspondingly. Much lower diffusion in the SPS tunnel can be explained by the comparatively low depth of the SPS and the relatively hard rock at the CERN cite. It has also been recently pointed out that if a long-term systematic motion is excluded then purely diffusive component of the SLAC linear accelerator tunnel motion exhibit much lower diffusion coefficient $A_{SLC} < 10\cdot10^{-6}$ $\mu m^2/s/m$ [30]. It has to be noted also, that for all the data presented in Fig.17 the exponent $\gamma$ of a power-law fit $<dY^2(L)> \propto L^{\gamma}$ varies between 0.7 and 1.0.

## 2.4 Geophysics Measurements Data on Ground Diffusion

Evidences of the ground diffusion either in space or in time or simultaneously in space and time have been reported in geophysics studies of various types. Below we present many of these results, classifying them by the method of the measurements: made with optical and laser interferometers, stretched wire and several types of HLSs.

*2.4.1 Strain Measurements in PFO*

Horizontal motion of massive near surface monuments emplaced in competent, weathered granite has been made by laser interferometers ("optical anchors") at Pinon Flat Observatory (PFO) in southern California [9]. The data on the optical path difference *dL* over the distance *L*=732 m have been normalized in the units of strain *ε=(dL/L)* and its power spectral density is shown in Fig.21 from [31]. The peaks in the spectrum around multiples of 1 cycle/day are caused by earth tides and temperature effects; the peak at high frequencies of ~0.1 Hz is caused by microseisms ("7-second hum"). Except for these peaks, the spectrum is very well fit by the power law $1/f^2$. Correspondingly, the rms wander $<(\varepsilon(t)-\varepsilon(t+T))^2>$ scales linearly with time *T* as demonstrated in the lower plot of Fig.21. From the linear slope, the ATL coefficient can be calculated as $A_{PFO} = <(\varepsilon(t)-\varepsilon(t+T))^2>L/T = 0.7\cdot10^{-6}$ μm²/s/m. The diffusion is very small compared to any examples we considered above – that is no surprise given that the PFO has been located in a very stable area with hard granite bed-rock suitable for very precise geophysics observations.

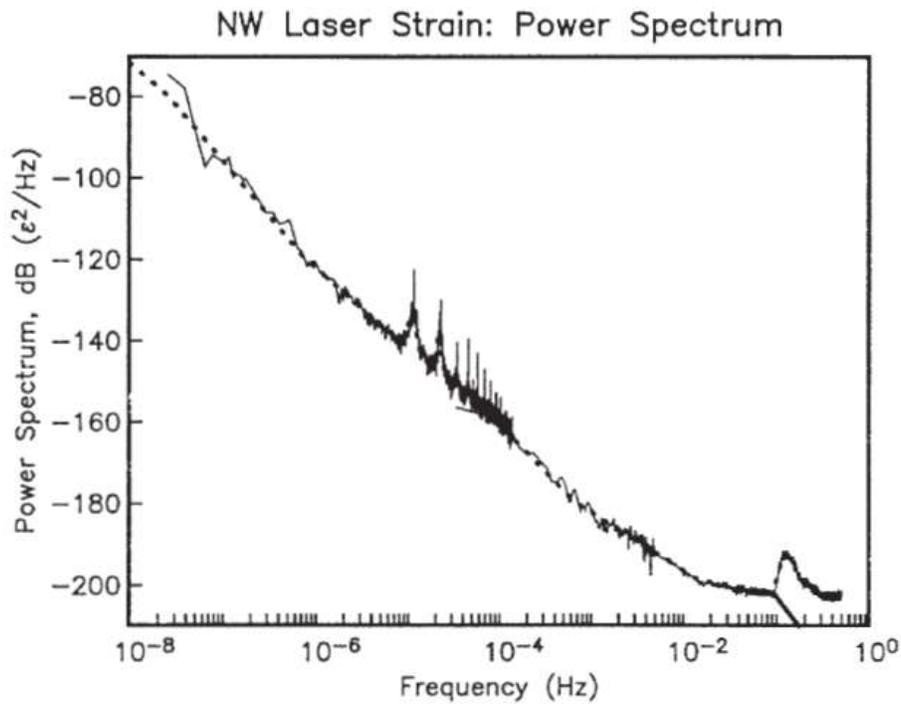

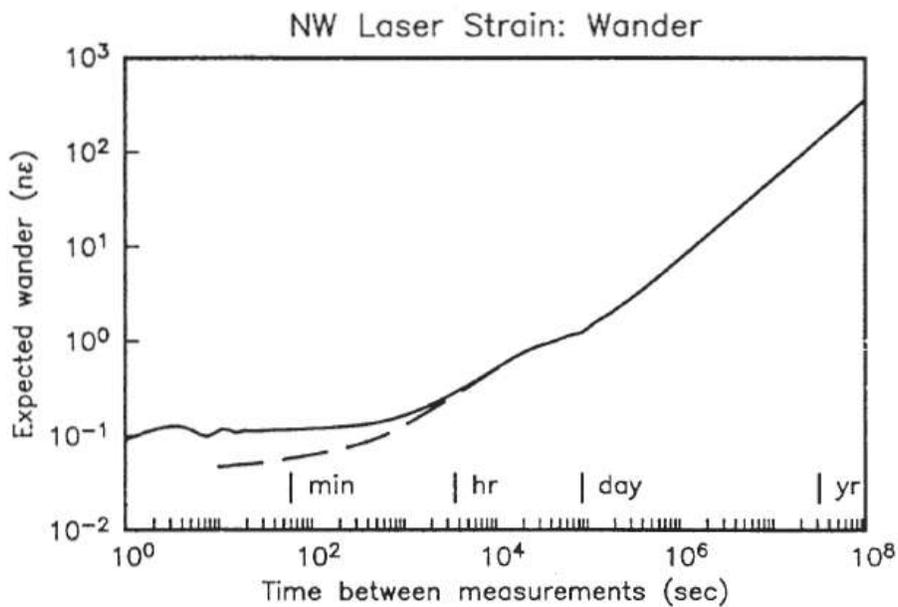

Fig.21 (top) PSD of the earth strain at Pino Flat Observatory in southern California; (bottom) the solid line is rms wander of the earth computed from the full spectrum, and the that computed if the "7-second" microseism peak is fileterd out, from [31].

*2.4.2 Laser Beam Measurements in the SLAC Tunnel*

Several measurements of slow ground motion were performed using laser alignment system [32] installed in the SLAC 2-mile linear accelerator tunnel. This system consists of a light source, a detector, and about 300 targets, one of which is located at each point to be aligned over a total length of 3050 m. The target is a rectangular Fresnel lens which has pneumatic actuators that allow each lens to be flipped in or out. The targets are installed in a 2-foot diameter aluminum pipe which is the basic support girder for the SLAC linear accelerator.

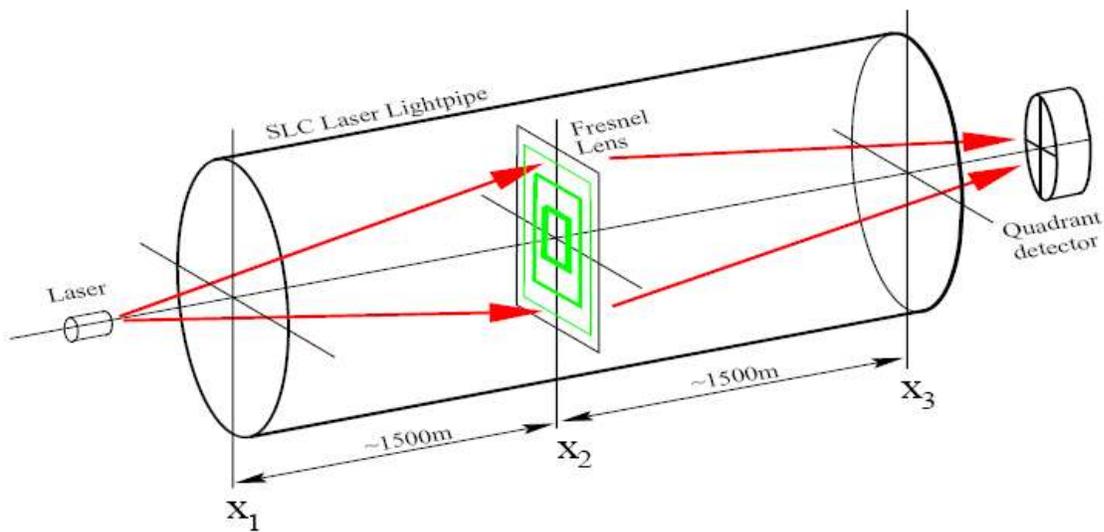

Fig.22. Schematic of SLAC linear accelerator laser measurement system.

The light source is a He-Ne laser shining through a pinhole diaphragm. The beam divergence is large enough to cover even nearby targets and only transverse position of the laser, but not angle, influences the image position. The light pipe is evacuated to about 15 microns of Hg to prevent deflection of the alignment image due to refraction in air. Sections of the light pipe, which are about 12 meters long, are connected via bellows that allow independent motion or adjustment. The measurements reported below were done with a single lens inserted which was not moved until the measurements were finished in order to ensure maximal accuracy. (In multi target mode the repeatability of the target positioning limits the accuracy). The schematic of the measurements with just one of the lenses exactly the middle of the system is shown in Fig.22. In such configuration, the laser spot position in the detector is equal to $x_1+x_3-2x_2$ (for either vertical or horizontal plane – see Fig.22).

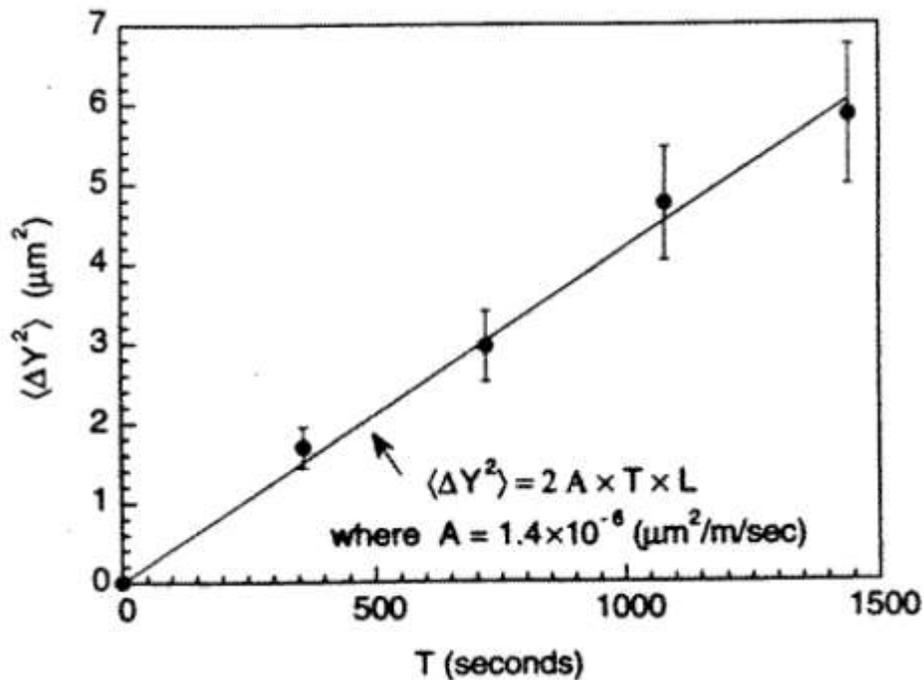

Fig.23: Variance of the vertical laser spot movement in the SLAC laser system (from Ref.[33]).

Analysis of the spot's vertical position variation shows that the variance of the motion scales linearly with time – see Fig.23 from Ref.[33] - that is consistent with $A_{SLAC} = 1.4 \cdot 10^{-6}$ μm$^2$/s/m.

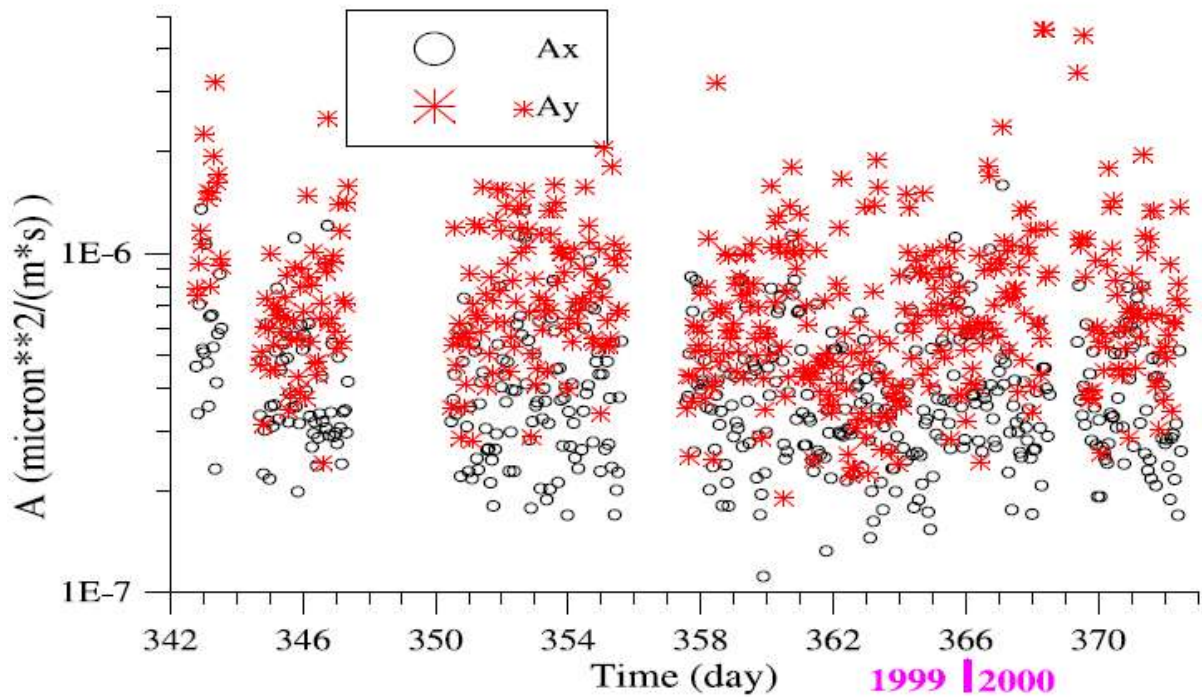

Fig.24. Diffusion coefficient $A$ as measured from the spectra laser spot vertical and horizontal movements in the frequency band 0.00024Hz to 0.015Hz (from Ref. [34]).

In the other series of measurements, reported in Ref.[34], it was found that the amplitudes diffusive motion in vertical and horizontal planes are about the same, see Fig. 24, and the excess in the vertical plane is often correlated with the atmospheric pressure variations.

### 2.4.3 Motion of the CERN PS Pillar

Yet another manifestation of the ground diffusion is the movement of central CERN Proton Synchrotron (PS) pillar over period of more than 2 years shown in Fig.25 from Ref.[35].

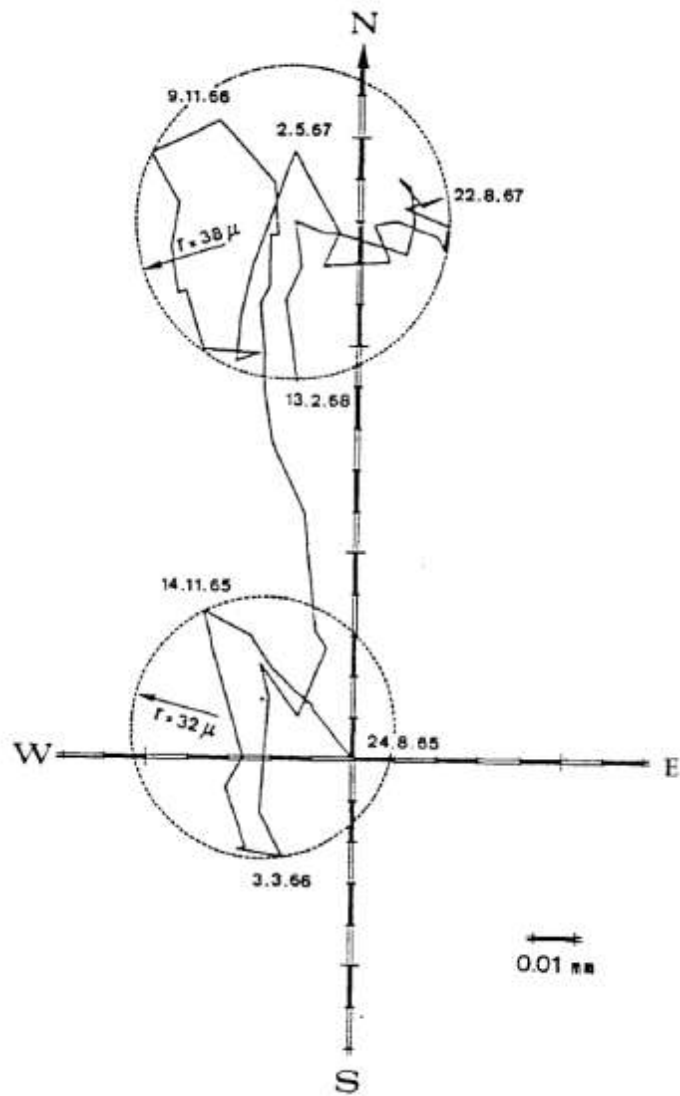

Fig.25:  Horizontal movement of the PS central pillar in 1965 - 1968 (from Ref.[35]).

A pair of horizontal pendulums was mounted on the PS pillar anchored in the molasses 10 m below ground level. These instruments measure the variations of their support in relation to the direction of the vertical, and, therefore, the movement of the vertical axis of the 10 m deep pillar. Such an inverted pendulum performed irregular motion that looks like Brownian motion. Extracting some linear trend (well remarkable in South-North direction), one can find, that in both directions the variance grows about linearly in time, and the coefficients of the *ATL* diffusion are equal to $A_{PS} = (3.0 \pm 1.0) \cdot 10^{-6}$ μm$^2$/s/m, that corresponds to the variance of displacement of about 500-900 μm$^2$ over the time interval of *T*=9 months and *L*=10m [36].

### 2.4.4 Stretched Wire Measurements at the SLAC FFTB Facility

A ~40 m long stretched wires were used for measurements of vertical and horizontal positions of several magnets in SLAC Final Focus Test Beam (FFTB) tunnel [37]. The magnets were divided into four sections with two parallel stretched wires in each section ("left" and "right" wires). The wire lengths vary from 30 m to about 43 m in the different wire sections. Each wire was stretched with a weight of about 35 kg at one end. Each magnet had submicron resolution wire position monitors attached to it.

The measurements were taken over about a week in the FFTB hall with a measurement point every 6 seconds. The hall has a thick concrete slab floor and was sealed to avoid thermal variations for most of the measurement interval. The results shown in Fig.26 indicate that the element positions wander in both vertical and horizontal planes with the diffusion coefficients in the range $A_{FFTB} = (4\pm3)\cdot10^{-6}$ $\mu m^2/s/m$.

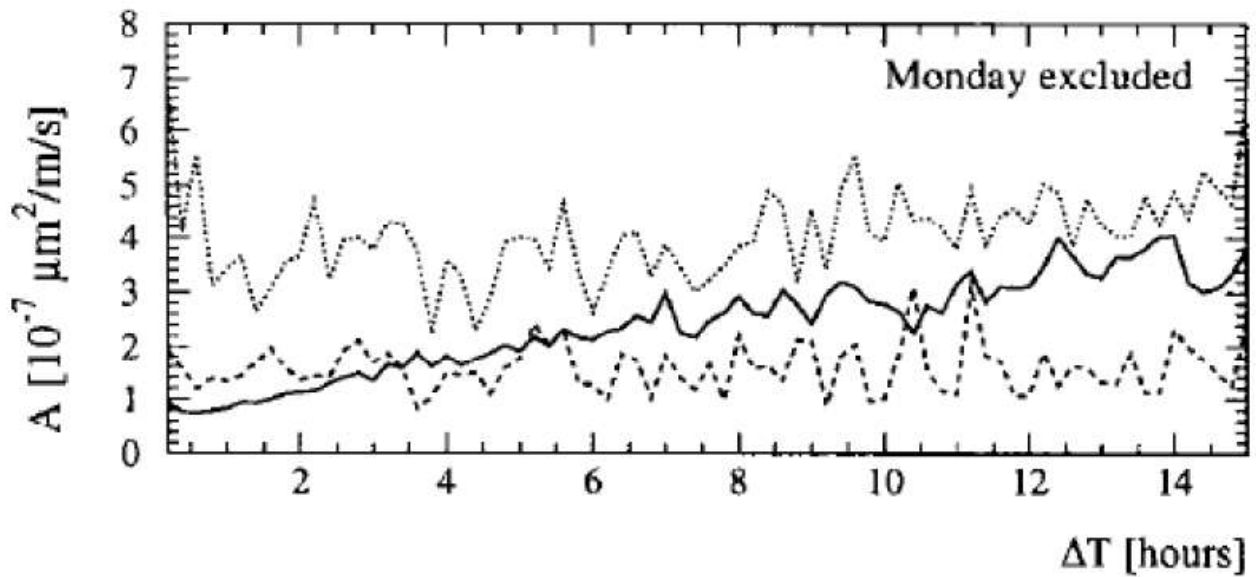

Fig.26: Calculated diffusion constant $A$ as a function of the time interval $\Delta T$ in the ATL rule. The three different curves refer to the horizontal (solid) and vertical (dotted) data of section 1 and the horizontal data of section 2 (dashed). The upper results include all data. In the lower case the some data were excluded from analysis to eliminate a perturbation effect of an FFTB access for one of the days was The $A$ constant was determined over a distance of about twice 15 m (from Ref. [37]).

*2.4.5 HLS Measurements in Japan*

Below we review several slow ground motion measurements made with HLS sensors made in various locations in Japan: in geophysics laboratories, in accelerator facilities and in several tunnels. More detail descriptions of the conditions and instruments can be found in the cited References.

*2.4.5.1* The Esashi Earth tide station is situated in the northwest of Japan. It occupies a tunnel in granite mountain side. Two $L$=50-m long water levels directed to South-North and East-West are at about 160 m from the tunnel entrance and about 60 m under the mountain surface. These tiltmeters detect vertical elevation difference. Observations started in June 1979 by National Astronomical Observatory Mizusawa. Fig.27 presents almost 15-years-long record of S-N and E-W tilts measured monthly [38]. Linear trends were extracted from the original data records and the variogram of the tilt $<d\Theta^2(T)>=<(\Theta(t)-\Theta(t+T))^2>$ calculated in [17]. The results are presented in Fig.28 and the data can be approximated by the linear fits of *0.026* µrad$^2$/month for the N-S tilt data and *0.018* µrad$^2$/month for the E-W tilt data (see dashed lines in Fig.28).

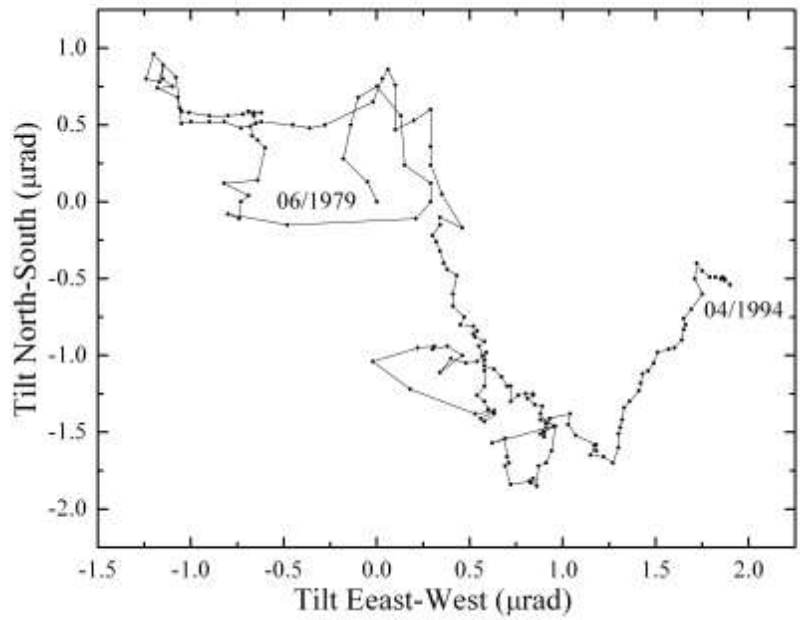

Fig.27:  Secular tilting motion measured at Esashi station in 1979-1994  (from Ref

[38], original data records courtesy of Prof. S.Takeda of KEK, Japan).

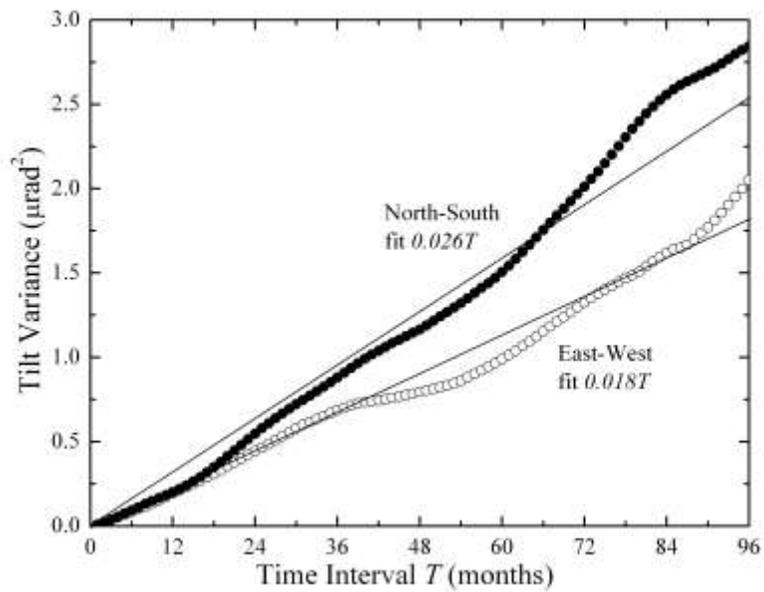

Fig.28:  Variance of the tilt elevation vs time interval (from Ref.[17]).

The observed time dependence of the variance $\propto T$ is a characteristics of a random walk (or Brownian) process. If one assumes the validity of the *ATL* law, than the diffusion coefficients can be estimated as $A_{ESNS} = <d\Theta^2(T)>L/T \approx 0.51 \cdot 10^{-6}$ $\mu m^2/s/m$ for the N-S tilt variations and $A_{ESEW} \approx 0.35 \cdot 10^{-6}$ $\mu m^2/s/m$ for the E-W tilt drifts.

*2.4.5.2* Series of high precision ground motion measurements with several hydrostatic level systems has been performed by the group of Prof. S.Takeda of KEK (Japan) since early 1990's. A 50 m long HLS system with an overall accuracy of 0.1 $\mu m$ was used in an old Sazare mine (Sumitomo Metal Mining Co., Ltd., Shikoku, Japan) located about 300 m under the surface of hard rock (green schist) mountain slope. The detected tilt was found to be a superposition diffusive of drifts, tides and precipitation effects – see the PSD of the tilt observed in a month long observations in 1993 in Fig.29 from [39]. One can clearly see several tidal peaks in the spectrum. The straight line indicates the $1/f^2$ dependence that corresponds to the *ATL* law spectrum Eq.(5) with $A_{Sazare}=0.12 \cdot 10^{-6}$ $\mu m^2/s/m$. Significant seasonal variations were reported, too, with the diffusive having maximum in December 1992 and minimum in March 1993.

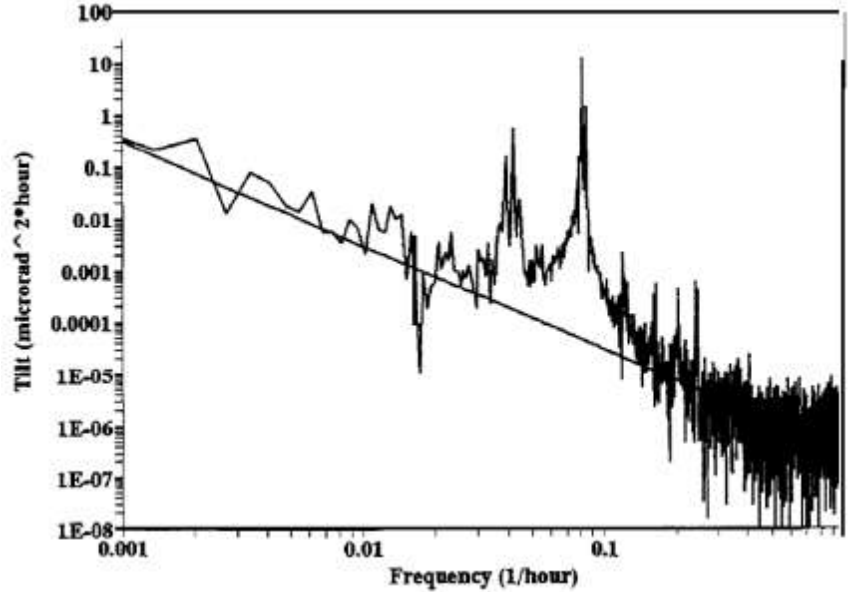

Fig.29: A spectrum of ground motion in Sazare mine (Japan). The straight line indicates $1/f^2$ (from Ref.[39]).

Similar studies with 12 m long and 42 m long water-tube HLS system were carried out in the tunnel of the TRISTAN storage ring (KEK, Tsukuba, Japan) and it was found that the power spectral densities could be also approximated by Eq.(5) with considerably bigger value of the diffusion coefficient $A_{TRISTAN\_HLS} \approx 40 \cdot 10^{-6}$ μm$^2$/s/m [40] – in a good agreement with the diffusion estimates obtained above from the TRISTAN orbit motion. It was noted, that the largest relative motion takes place across the different tunnel blocks separated by expansion joints.

The diffusion studies in several more tunnels in Japan confirmed that the ATL-law scaling Eq.(5) offers a very good fit to most of the data, and concluded that the diffusion parameter *A* is influenced dominantly by the earth and rock properties [41,42]. The observed parameter *A* is smaller in the tunnel in a solid rock than in the broken rock. The excavation method of the tunnel also affects significantly the diffusion: e.g. , a tunnel made by dynamite blasting had *A=5·10$^{-6}$* μm$^2$/s/m  while a tunnel in a similar rock bored by a tunnel-boring-machine had *A=1·10$^{-6}$* μm$^2$/s/m. Such a difference was attributed to artificial fragmentation of the rock occurred during the construction. Values of the diffusion coefficients measured in various Japanese tunnels will be presented in Table 1 below.

*2.4.6  HLS Measurements in Luxembourg*

Yet another example of the power-law ground drifts is measurements with a  43  m long floatless water-tube tiltmeter which has been in operation since 1997 at the Walferdange Underground Laboratory for Geodynamics in the Grand Duchy of Luxembourg [43]. The instrument 's very low noise level and its high resolution up to the long-period seismic band (where for instance the resolution is better than $5 \times 10^{-12}$ rad) allow the successful recording of miniscular drifts as well as rarely observed grave toroidal and spheroidal free oscillations of the Earth excited by major earthquakes. In the environmental conditions of its installation (in a gypsum mine at 100  m depth), the instrument shows a high degree of reliability and a very low drift rate (<0.005 microrad/month). The observed spectrum of the tilt is shown in Fig. 30 and has distinct power-law scaling at frequencies below 0.0001 Hz $PSD \propto 1/f^{2.2}$ (red dots); effective ATL diffusion constant at the lowest frequency of $f=2 \cdot 10^{-7}$ Hz can be found from Eq.(5) to be about $A=0.1 \cdot 10^{-6}$ μm$^2$/s/m.

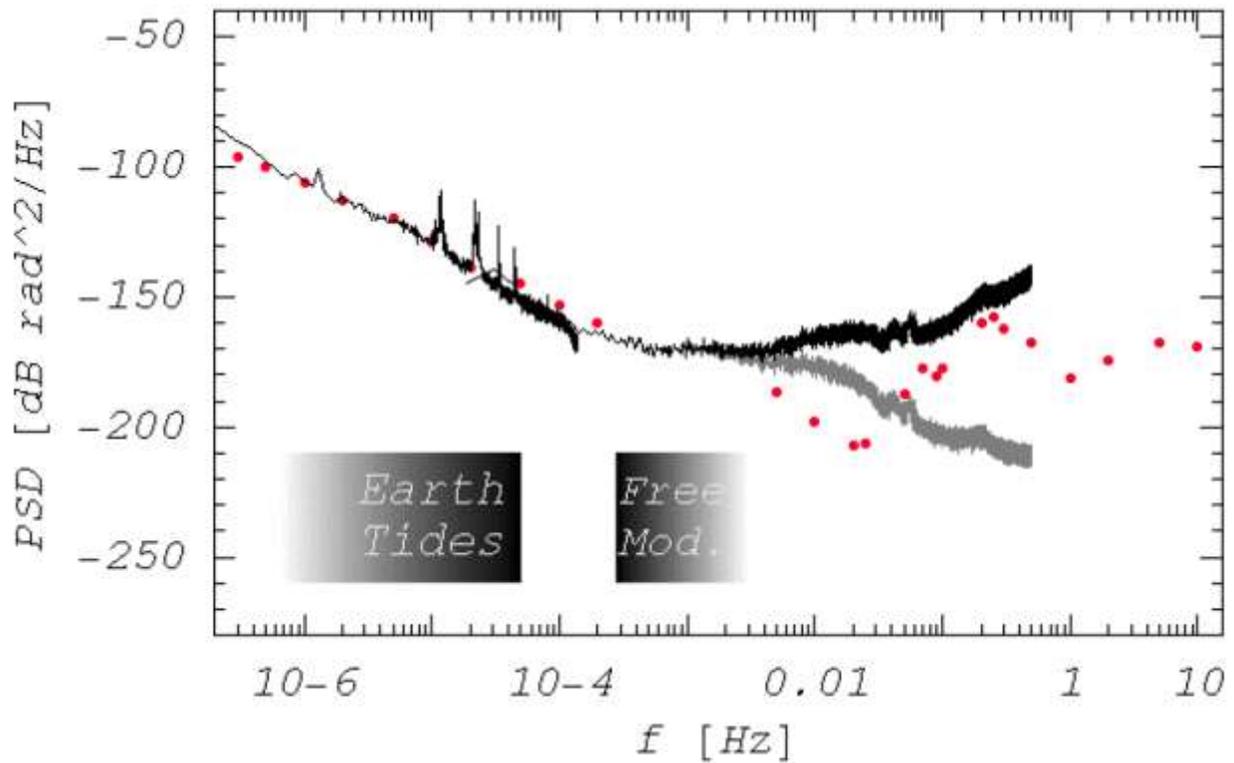

Fig. 30. The PSD of the 6 years long record in Walferdange (with and without the instrumental response correction - black and gray curve respectively) and the low tilt noise reference model from Ref. [9] (red dots). Shaded rectangles pinpoint the frequencies ranges of the Earth tides and the Earth free modes (from Ref.[43]).

## 2.4.7 Measurements with Multi-probe HLS systems in Illinois

The examples of the HLS measurements considered above provide confirmation of the ground diffusion in time as in all of them only two HLSs were used. To study the diffusion in space or spatial correlations of the ground motion, a series of extensive studies with systems of connected HLS probes has been performed in various locations in Illinois. High precision HLS probes developed for these studies (see Fig.31) are capacitive sensors equipped with local water temperature meters needed for thermal expansion compensation. The probes are made in two configurations – one for use with a single 1" diameter half-filled water pipe, and another for use with two separate ½" diameter tubes for air and for water (fully filled).

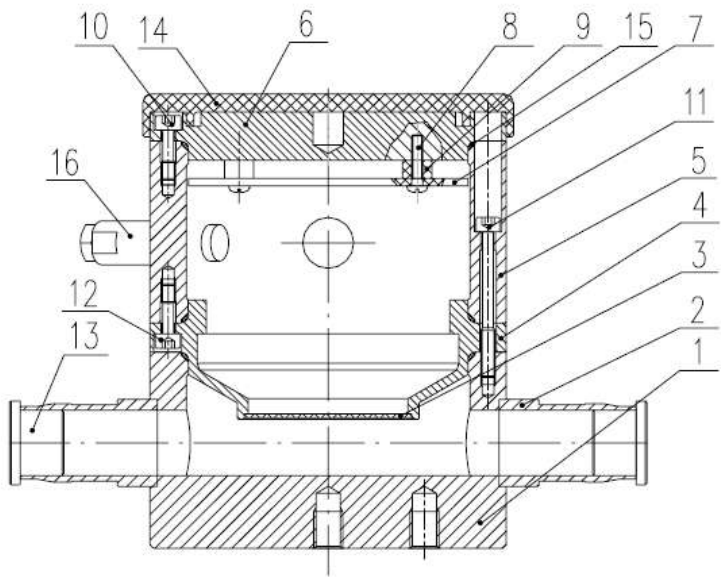 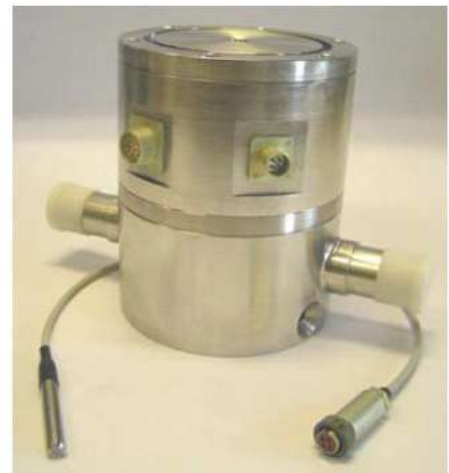

Fig.31: SAS-2 HLS sensor used in ground motion studies in Illinois (from Ref.[44]).

A pair of the probes set side-by-side shows the differential noise level of $\sigma^2 = (0.09\mu m)^2 + 1.252 \cdot 10^{-7} \ \mu m^2/s \cdot T$ (more details can be found in Ref.[44]). In a typical measurement arrangement, six to 20 of such probes installed in the same water level system spaced 15 to 30 meters apart usually along the line as shown in Fig.32. Once a minute, a PC based data acquisition system collects not only the water level data (averaged over the minute), but also all probe's temperature readings for correction, readings from one or two air pressure sensor for monitoring.

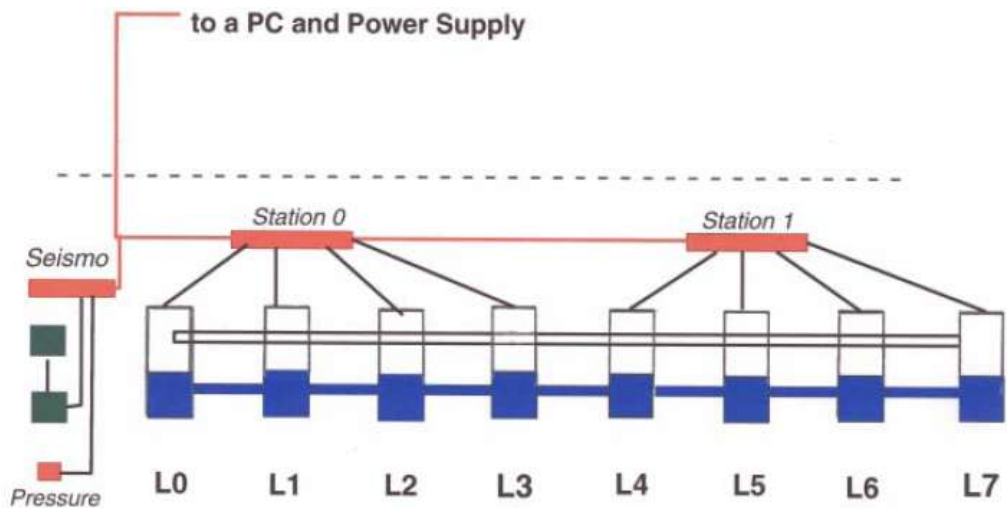

Fig.32: Schematics of the systems of HLS sensor used in the studies in Illinois.

*2.4.7.1* Studies in the Proton West (PW) tunnel on site of the Fermi National Accelerator laboratory had been carried out in 1999-2000 [45]. This is an unused beam line for fixed target experiments with a shallow (5 m depth) tunnel built by "cut-and-cover" method in 1970's. It has flat concrete floor that made quite easy the installation of 6 HLSs over total length of 180 m (30+30+60+30+30 meter apart).

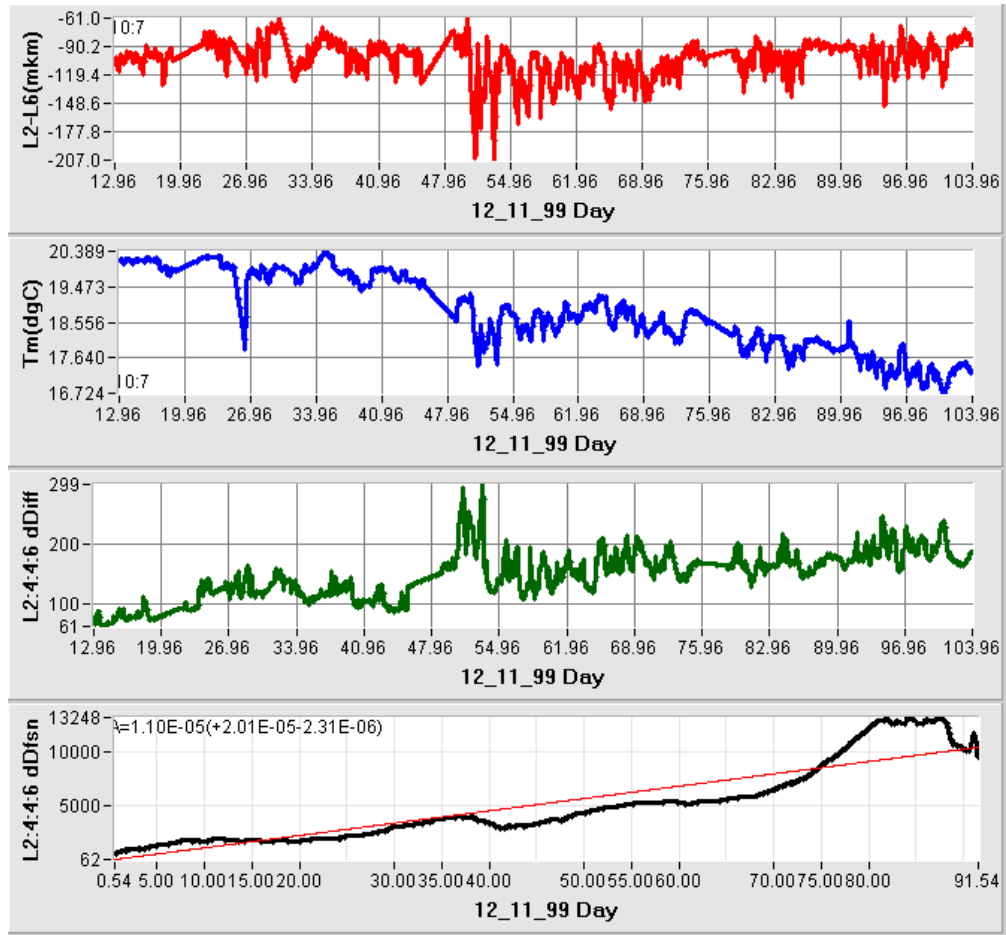

Fig. 33: 91 days data records starting November 12, 1999 from the PW studies: (top to bottom) the level difference between probes #2 and #6 120 (vertical scale of about 150 μm), mean temperature in the tunnel (vertical scale of 3.5 degree C), the second level difference $SD_{2446}$ (see in the text, scale 240 μm), and variance of the second level difference $SD_{2446}$ for intervals of up to 91 days (from Ref. [45]).

An important drawback of the tunnel was that it was not sealed and there were large temperature variations from one end to the other sometimes by few $^{o}$C a day causing large of changes in the water level readings – see in Fig.33. The ground tilts due to earth tides occured twice times a day with some 20 μm peak-to-peak amplitude in the level difference $Y_2 - Y_6$ between two probes #2 and #6 150 m apart but practically absent in the second difference $SD_{2446} = Y_2 - 2Y_4 + Y_6$. The variance of the second difference grows approximately linear with time interval $<SD_{2446}^2(T)> \approx T \cdot 114$ μm$^2$/day (see dashed line in the bottom plot in Fig.33). Making statistical analysis for all possible combination of probes one got the ATL law diffusion coefficient of about $A_{PW} = (6.4 \pm 3.6) \cdot 10^{-6}$ μm$^2$/s/m. The lack of data points in spatial intervals does not allow to confirm or reject the $L$-dependence of the proposed ATL model.

*2.4.7.2* Ground motion studies in the MI8 (Main Injector 8 GeV) tunnel took place over few months 2002-2003 and employed 20 HLS sensors equidistantly installed over 285 m long line (so, the probe-to-probe distance was 15 m) [46]. The tunnel is shallow and of a similar construction type and geology as the PW tunnel and the Tevatron tunnel discussed above. For several months the observed water levels data were dominated by a quasiperiodic motion with amplitude of about 10 *μm* every ~2 hours. Finally, the source was tracked to a domestic water well located 219 ft deep and several hundred feet away from the MI8 tunnel which slowly and periodically change ground water level. At the end, only one month of February 2003 was available for low-noise measurements of the ground diffusion. The coefficients $A$ calculated as $A=<SD_{nmml}^2(T)>/T/2L$ where the indexes $(n,m,l)$ indicate triples of the sensors distanced by $L$ and $T$=1 month are shown in Fig.34. E.g. the circles at $L$=120m data are for three combinations of the sensors (#1,#9,#17), (#2,#10,#18), (#3,#11,#19). One can see that the range of the $A$'s covers the PW results and roughly constant for distances $L$ from 15 m to 90 m. However, the mean value of $A_{MI8}$ =$(1-10)\cdot 10^{-6}$ $\mu m^2$/s/m appears to decrease with $L$, as if the variance scales as $dY^2 \propto T^l L^\gamma$ with $0<\gamma<1$).

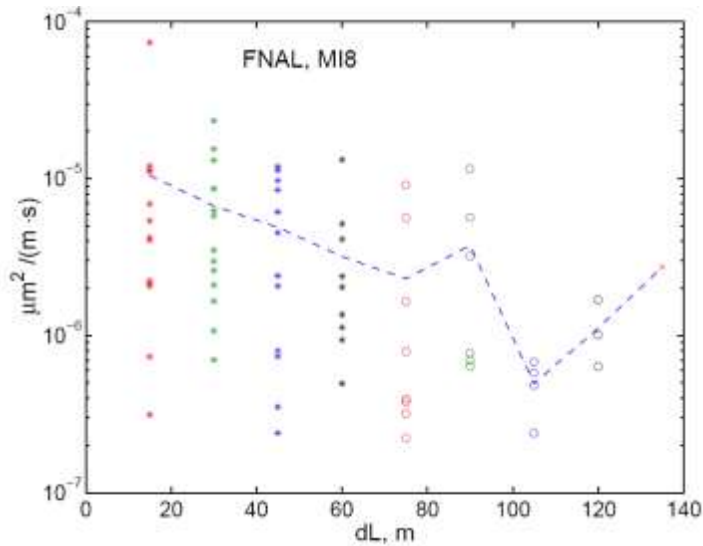

Fig. 34: Diffusion coefficient A calculated for all possible combination of the probes distanced by L from 15 m to 135 m from 1 month data records in MI8 tunnel [46].

*2.4.7.3* Since early 2004, a system of 20 HLS sensors with half-filled water pipe was installed in the Tevatron tunnel on top of the accelerator focusing magnets spaced 30 m apart – see Fig.35.

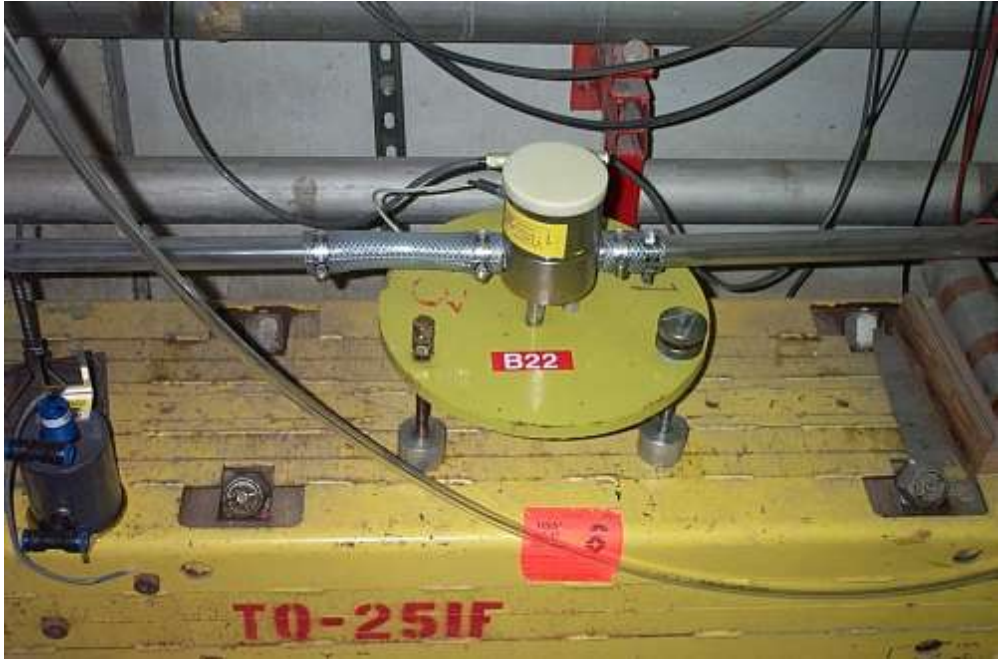

Fig.35. HLS probe on Tevatron accelerator focusing magnet.

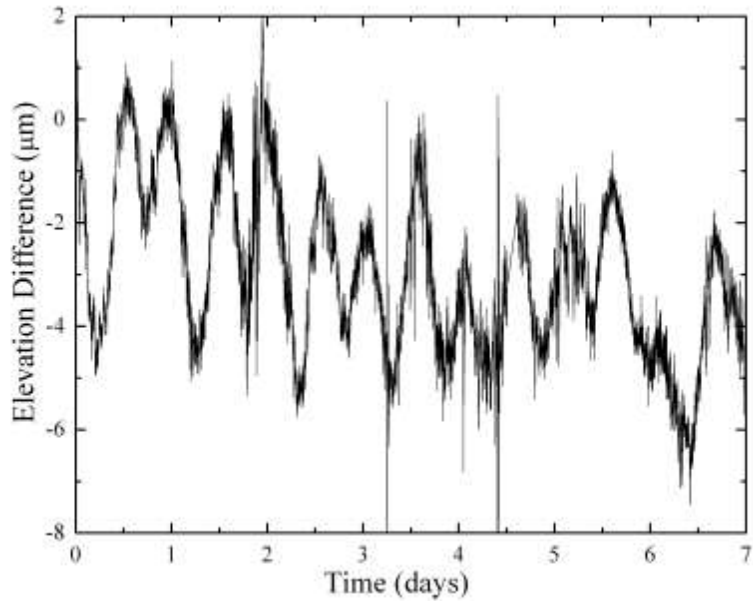

Fig.36. One week record of elevation difference of two neighbor focusing magnets in the Teveatron tunnel as measured by HLS (starts midnight Feb.7,2004; Ref.[29]).

Environment of a working accelerator had its own peculiarities, e.g. regular ramping of the electromagnets resulted in few micron relative magnet position changes – see spikes in Fig.36 from Ref.[29]– on top of regular tidal variations and diffusive drifts. Fig.37 shows a snapshot of the magnet elevation changes after 23 days of observations. One can see that the differential movements over the ~600 m section of tunnel could be as big as 30-50 μm.

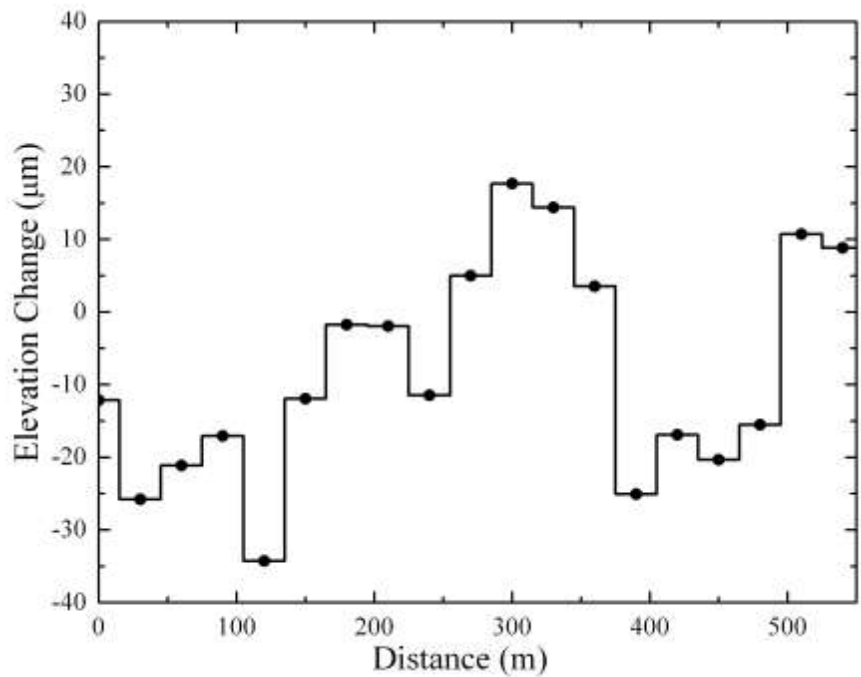

Fig.37. Change of the elevations of 20 Tevatron magnets after 23 days of observations  (Jan.7-Feb.1,2004; from Ref.[29]).

Variograms of the second differences have been analyzed, linear dependence on the time interval $T$ confirmed and the variance $<SD_{nmml}^2(T)>/T$ are plotted in Fig.38. As in the MI8 tunnel data analysis, the indexes $(n,m,l)$ indicate triples of the sensors distanced by $L$ and $T$=7 days – the week of Feb. 7, 2004. One can see that the variance increases with L up to 90-120 m and then flattens out. That indicates lack of coherence (independence) of the motion of the pieces of the tunnel distanced by more than 120 m apart – at the time scale of 1 week. For shorter distances, the ATL law with coefficient $A_{TevB}$ =(2.2±1.2)·10⁻⁶ µm²/s/m gives a good approximation of the data, in a good agreement with the diffusion estimate from the accelerator beam orbit motion discussed above.

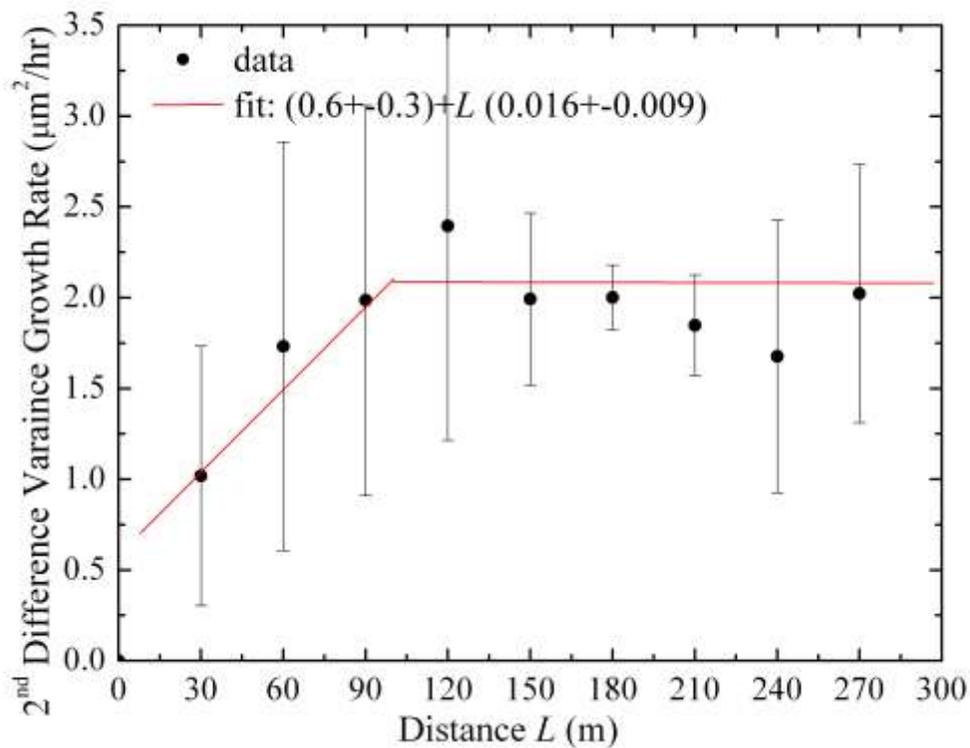

Fig.38. Dependence of the growth rate of the variance of the 2[nd] difference vs distance between the HLS probes in the Tevatron tunnel, the week of Feb 7,2004 (from Ref.[29]).

*2.4.7.4*   Seven HLS probes had been installed in 2006 in the MINOS experiment underground hall some 100 meters below grade on top of  the Galena Platteville dolomite (also on the site of Fermilab). The probes are set 30 m apart and connected in two double-pipe (air/water) systems – the first one with 4 probes  are orientated along a North-South line and the other system of 3 oriented  along an East-West line. One month long record of the HLS readings of the level difference $Y_0 - Y_3$  (probes #0 and #3, 90 m apart in NS direction) is presented in Fig.39.  One can see that some 6 μm amplitude  periodic variations due to the Earth tide  dominate few μm scale slow drifts over weeks.

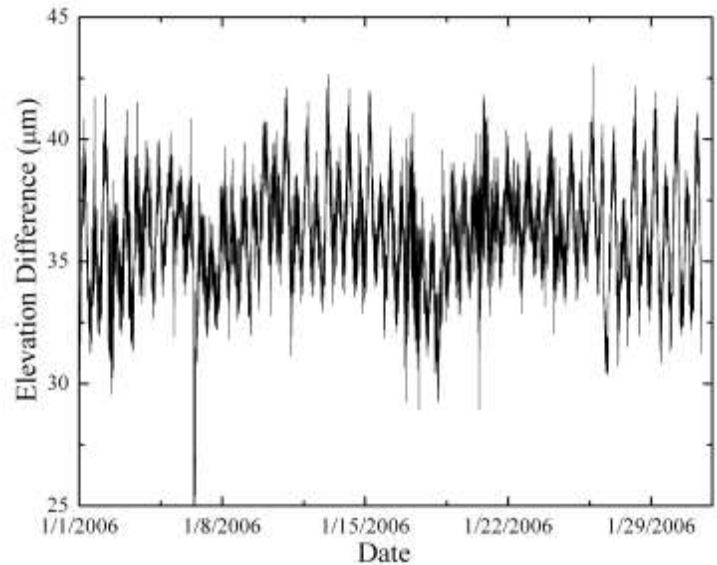

Fig.39. January 2006 record of elevation difference for two HLS probes 90 m apart

in the FNAL MINOS hall [29].

To remove the systematic effects due to the tides, the FFT of the 1 month long record of the level difference $Y_0 - Y_3$ data has been calculated (see Fig.40). The power law fit *1/f* indicated by the red line in Fig.40 corresponds to the ATL diffusion coefficient of $A_{MINOS} = 0.18 \cdot 10^{-6}$ $\mu$m$^2$/s/m [29].

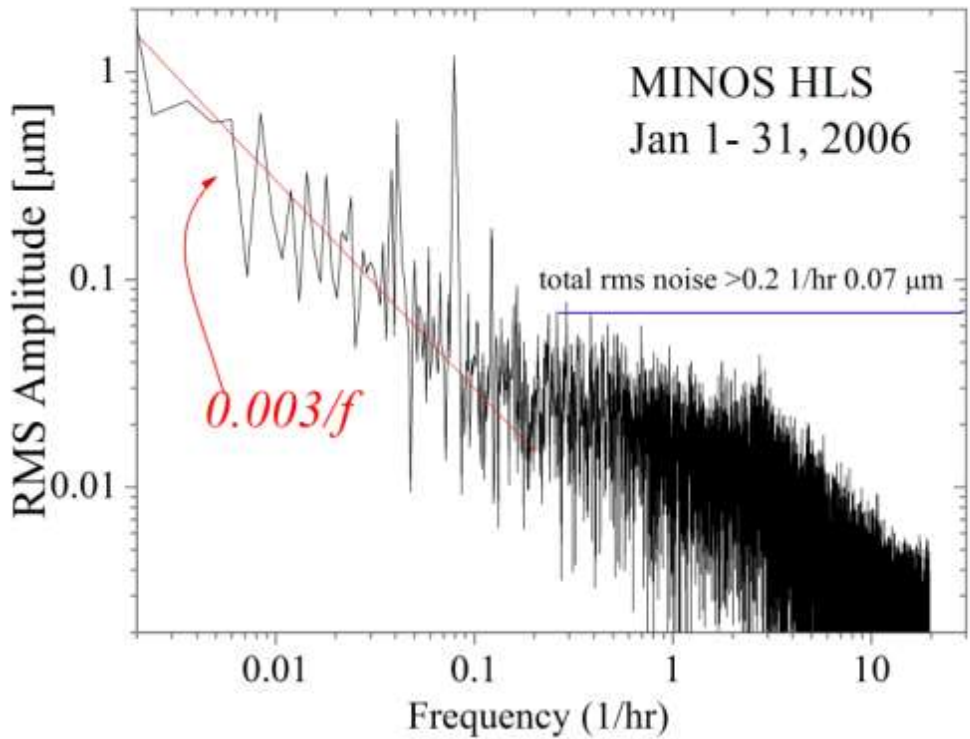

Fig.40. FFT of the elevation difference for HLS probes 90 m apart as measured in the Fermilab's MINOS hall [29].

*2.4.7.5* Since early 2000, continuous slow ground motion measurements with up to 8 HLS probes are being carried out in a 100 m deep dolomite mine (Conco-Western Co./LaFarge Co. , North Aurora, IL) – some 3 miles South-West of Fermilab.

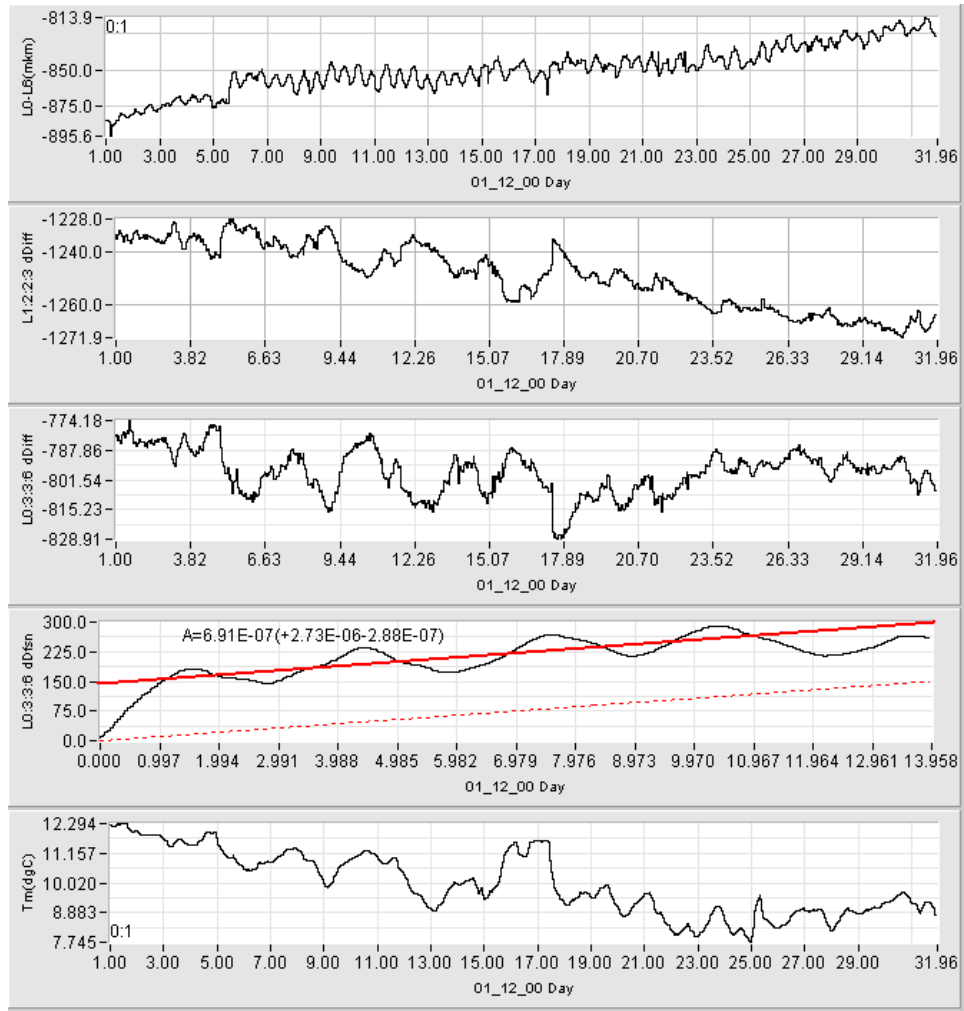

Fig. 41: Slow ground motion in 120 m deep dolomite mine (Aurora, IL) in

December, 2000. Top to bottom a) to e), see comments in the text [47].

This is a multi-layer mine in Galena-Plattville dolomite. Our 210 m long system was set at the depth of about 80 m near the border wall of this 0.8km×1.4km underground facility. During the studies the mine continued dolomite production and some 3 tons of explosives were detonated each day at around 3 p.m. except weekends in different areas and at different levels of the mine. Ventilation system makes the temperature of mine very dependent on the outside temperature. Fig.41 shows one month data records in the Aurora mine in January 2000. The horizontal axis is time in days in December 2000 (e.g., 31.96 corresponds to late night of December 31, 2000). The vertical axis on the Fig.41 a) is for a relative vertical position of two observation points 180 meters apart (total scale is 895-813=82 µm). Because of periodic changes in relative positions in the system Moon-Earth-Sun, the amplitude of diurnal oscillations varies with a period of 14 days – it is obviously less at the beginning of the plot and in the middle of the month. Obvious creep (slow change of the tilt) of the order of 82 µm/180 meters=0.5 µrad is seen over 1 month in the same plot. Possible explanations for this change are: natural geological instability, temperature effect or atmospheric pressure effect. Fig.41 e) reveals 1 $^{\circ}$C variations in the Aurora mine daily and some 4 $^{\circ}$C drop in the temperature over 3 weeks. To separate the temperature effects and the ground diffusion from the tides, the second difference $SD_{1223}$ for the probes 30 m apart $SD_{0336}$ for the probes 90 m apart are computed and plotted in Fig.41 b) and c). One can see that they are correlated with the average temperature changes with coefficients about $-20$ µm / $^{\circ}$C and $+40$ µm / $^{\circ}$C correspondingly. Air pressure also can contribute into the motion of the ground, both in $SD_{1223}$ and $SD_{0336}$ but it is usually prominent only over longer distances of $L>1$ km. Besides regular Earth tides and temperature drifts, the ground does move randomly due to the natural diffusion. Fig.41 d) shows the mean square of the second vertical difference for the points 90 meters apart, and the red line presents linear fit $<SD_{0336}^2(T)> = 150 + 2\,ATL$, with $A=0.69 \cdot 10^{-6}$ $\mu m^2/s/m$ and $T$ up to 14 days. Somewhat excessive motion at short periods $T<1$ day can be explained by the ground jumps due to the daily blasts taking place in the mine (within 1 mile from the measurement system location) – several of them with amplitudes of 10 to 25 microns are seen in the Fig.41 a). Extraction of temperature correlated signals and linear drifts leads to the average (over all combination of the second differences and over all possible $L=30, 60, 90$ m) value of $A_{Aurora} =(0.58 \pm 0.28) \cdot 10^{-6}$ $\mu m^2/s/m$.

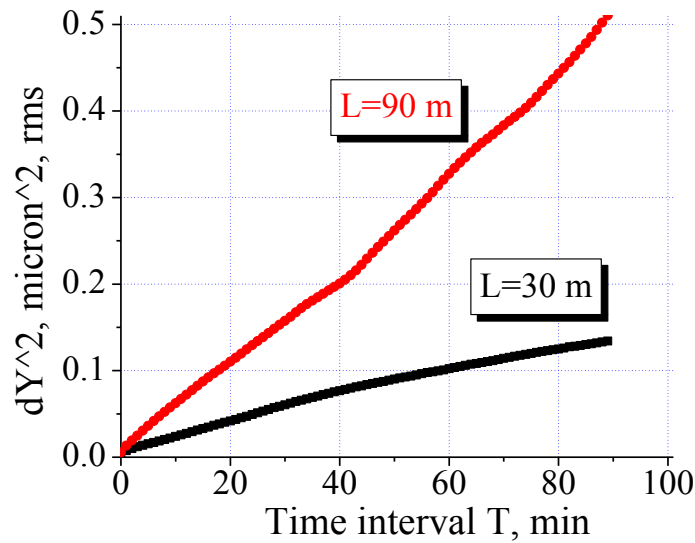

Figure 42: Variance of the vertical relative ground motion for the points 30 m and 90 m apart, measured on October 13-15, 2000 in Aurora mine, IL [46].

There were no blasts over weekends as well as sometimes the temperature does not change much as well, so one can use such records for analyzing "natural" ground diffusion at shorter time scales. For example, on a quiet weekend of Oct. 13-15, 2000, the temperature variation was less than $0.05^{o}$C. The 2 days record analysis is presented in Fig.42 which shows the variance of the second differences $<SD_{1447}^{2}(T)>$ ($L=$ 90m, red circles) and $<SD_{1223}^{2}(T)>$ ($L=$30 m, black squares) for the time intervals of up to $T=$90 minutes. In a good accordance with the ATL law, the variances grow linearly with $T$, the variance is about 3 times larger for a 3 time larger distance, and corresponding diffusion coefficients are almost the same $A_{A90}$ $=0.53\cdot10^{-6}$ $\mu$m$^{2}$/s/m and $A_{A30}$ $=0.42\cdot10^{-6}$ $\mu$m$^{2}$/s/m.

## 3 Discussion on Fractal Nature of the Ground Diffusion

### 3.1 Discussion of the results

Several conclusions can be made from the results presented above. First of all, the diffusive motion of the ground is often just a background to much more powerful processes, like ground expansion due to temperature changes, or bending due to atmospheric pressure variation or winds, long-term settlement drifts or Earth tides. Special data processing is often needed to separate diffusive noise from systematic or periodic signals: in time- or space- domains, that can be achieved with use of digital filters, like *the first* or *the second difference* methods employed above; in the

frequency- or wavelength- domains, Fourier analysis of windowed data sets (e.g. with Hanning window) makes visible the power-law component of the spectrum.

Table I below summarizes the observations of the ground diffusion presented above and presents the diffusion coefficient $A$, time interval $T$ of the observation or analysis, the spatial scale $L$ (e.g. the tunnel length, of the total length of the HLS system), plane (V is for vertical, H is for horizontal) and effective depth of observations. The second column indicates whether temporal ($T$) or spatial ($L$) characteristics of the diffusive ground motion have been explored. One can see that most of the accelerator orbit drift data and most of the HLS and laser interferometer studies reveal the diffusion in time. Many accelerator alignment data manifest the diffusion in space. Diffusion in both time and space is observed in many-year accelerator alignment data and in long-term measurements with HLS systems employing many (up to 20) probes.

TABLE I. Summary of Ground Diffusion Measurements

| | | $A$, $10^{-6}$ $\mu m^2/s/m$ | Time | Scale | Ref. | Comments |
|---|---|---|---|---|---|---|
| | | *Beam Orbit Drifts in Accelerators* | | | | |
| HERA -$e$ Vertical | $T$ | 4±2 | 25 days | 6.3 km | [18] | 25m deep, $\Delta L$=23m |
| HERA -$p$ Vertical | $T$ | 8±4 | 5 days | 6.3 km | [18], [17] | 25m deep, $\Delta L$=47m |
| TRISTAN Vertical | $T$ | 27±7 | 2 days | 3.0 km | [19], [17] | 12m deep, $\Delta L$=47m |
| Circumf. KEK-B | $T$ | 27±3 | 4 months | 3.0 km | [20] | $\beta \approx 2.2$ |
| Tevatron Vertical | $T$ | 2.6±0.3 | 15 hrs | 6.3 km | [21] | ~7m deep, $\Delta L$=30m |

| | | | | | | | |
|---|---|---|---|---|---|---|---|
| | Horiz. | $T$ | 1.8±0.2 | 15 hrs | 6.3 km | | |
| LEP | Vertical | $T$ | 10.9±6.8 | 18 hrs | 26.7km | [22], [24] | ~100m deep, $\Delta L$=39m |
| LEP | Vertical | $T$ | 39±23 | 3.3 hrs | 26.7km | [23] | tides not excluded |
| | Horiz. | $T$ | 32±19 | 3.3 hrs | 26.7km | [23] | tides not excluded |
| SPS | Vertical | $T$ | 6.3±3.0 | 2 hr | 6.9 km | [23] | 50m deep, $\Delta L$=32m |
| *Accelerator Alignment Data Analysis* | | | | | | | |
| CERN LEP Vert | | $L,T$ | 6.8-9.0 | 6, 9 mos | 26.7 km | [26],[17] | 45-170 m deep |
| | | | 3±0.6 | 6 years | 26.7 km | [27] | $\Delta L$=39m |
| CERN SPS Vert. | | $L,T$ | 14±5 | 3-12 yr | 6.9 km | [28] | 50 m deep, $\Delta L$=32m |
| Tevatron Vertical | | $L,T$ | 4.9±0.1 | 1-6 yr | 6.3 km | [29] | ~7m deep, $\Delta L$=30m |
| SLAC PEP Vert. | | $L$ | 100±50 | 20 mos | 2 km | [28] | cut-and-cover tunnel |
| SLAC Linac Vert | | $L$ | 200±100 | 17 yr | 3 km | [15],[28] | cut-and-cover tunnel |
| | | $L$ | <10 | 17 yr | 3 km | [29] | linear trends removed |
| UNK Site Vert | | $L$ | 100±50 | 2 yr | 500 m | [15],[28] | surface monuments |
| *Geophysics Instruments Data* | | | | | | | |
| PFO (CA, USA) | | $T$ | 0.7 | 5 year | 732 m | [31] | laser interferometer |
| SLAC Linac Vert | | $T$ | 1.4±0.2 | 0.5 hr | 3 km | [33] | $\Delta L$=1500m |
| | | $T$ | 0.2-2 | 1 hr | 3 km | [34] | from PSD fit |
| CERN PS pillar | | $T$ | 3±1 | 2.5 yr | 10 m | [35],[36] | 10 m depth |
| SLAC FFTB | | $T$ | 0.1-0.5 | 15 hrs | ~30 m | [37] | wire, in the lab |
| Esashi (Japan) | | $T$ | 0.3-0.5 | 15 years | 50 m | [38],[17] | 60m deep, NS-EW |
| Sazare (Japan) | | $T$ | 0.01-0.12 | 6 weeks | 48 m | [39] | 300m deep |
| Kamaishi (Japan) | | $T$ | 0.06-0.14 | | | [42] | Granite |
| SPring8 (Japan) | | $T$ | 0.8 | | | [42] | Granite |
| Miyazaki (Japan) | | $T$ | 15 | | | [42] | Diorite |
| Rokkoh (Japan) | | $T$ | ~36 | | | [42] | Granite |
| KEKB tunnel | | $T$ | 40 | 4 days | 42 m | [40] | 12m deep, joints |
| FNAL PW7 | | $T$ | 6.4±3.6 | 3 months | 180 m | [45] | $\Delta L$=30m, $t^o$-effects |
| FNAL MI8 line | | $T,L$ | 1-10 | 1 month | 285m | [46] | $\Delta L$=15m, m.b. $\gamma<1$ |
| FNAL Tevatron | | $T.L$ | 2.2±1.2 | 1 week | 600m | [29] | $\Delta L$=30m,$\gamma=0$ $L>120m$ |
| FNAL MINOS hall | | $T,L$ | 0.18 | 1 month | 90m | [29] | $\Delta L$=30m, ~100 m deep |
| Aurora mine (IL) | | $T,L$ | 0.6±0.3 | 2 weeks | 210m | [46],[47] | $\Delta L$=30m, ~100 m deep |

Another conclusion which can be made is that the speed of the diffusion - the coefficient $A$ – is site dependent and has tendency of being smaller at bigger depths, in harder rocks and in geologically stable locations (like like those where geophysical observatories are set at). Japanese data indicate that even the tunneling method may affect the diffusion rate.

One can also see that the *ATL* approximation is not always the best, and in general, the exponents in the fit $<dY^2(T,L)> \propto T^\alpha L^\gamma$ can significantly differ from 1. It has to be noted that some of our observations show that at small time intervals $T$ and large spatial separations $L_m$, the motion of two points naturally independent and, therefore, the exponent $\gamma$ tends to be close to 0.

With the limited number of data sets, we can not explore in detail the boundary $L_m(T)$ beyond which the independence (or significant loss of correlation) occurs while it is a very important phenomena [48] which definitely calls for more studies.

The observations reviewed above cover time intervals from hours to several years and spatial scales from dozen meters to a dozen of kilometers (the largest accelerators). There are some evidences of the diffusion at much larger $T$ or $L$ intervals. For example, 50 years observation (1930-1980) of sea levels in 12 Japanese ports distanced by as much as 800 km [49], showed that besides daily and seasonal changes, the level variation has a long-term "random walk" component $<dY^2(T)> \propto T$ with computed diffusion coefficient $A$ of about $35 \cdot 10^{-6}$ $\mu m^2/s/m$ [17]. It is long known to geophysicists, that Earth's topography is fractal, and its power spectral density scales with the wave number as $S(k) \propto k^{-2}$ that corresponds to $<dY^2(L)> \propto L$ over distances 100 km to 6000 km (see, e.g. Fig.17.19 in Ref. [14] and corresponding discussion). What this paper adds to previously known results is the notion that the diffusion takes place *both* in time and in space (at least, over the scales indicated in the Table I and characteristic for high energy physics accelerators).

### 3.2 Modeling Diffusive Ground Motion

The fractal objects and time series are one of the favorite subjects for modern studies on geophysics, geomorphology, hydrology, landscape evolution, etc, and a variety of models have been proposed and studied in great detail (see e.g. [12,13,14,50] and references therein). To simulate the "ATL law" in computer codes for accelerator design, several algorithms that produce the required space and time dependencies have been developed. In the case of a linear system (points of the ground are equally distributed along a straight line) it could be a straightforward to apply the "random walk" procedure: for a given time step $k$ it is only necessary to start at one end, giving each point a random displacement $\Delta_m^k$ with respect to the previous point $Y_i^k = Y_i^{k-1} + \Sigma_{m=0}^i \Delta_m^k$ [6, 18]. It is easy to see that the variance of resulting relative displacement of any two points separated by $L$ is given by *ATL-law* Eq.(4). With a bit more cumbersome mathematics, the method can be extended to any one dimensional geometry shape (e.g. circle) on a two dimensional surface [51].

Below we present a simple one-dimensional model of the landscape evolution which has certain physical meaning, satisfies the ATL-law, and reveals a reduced correlation of the surface motion at large distances. The model considers the ground as a set of separated blocks with different characteristic sizes $R$ – as approximately shown in Fig.43. The number of blocks $N_b(L,R)$ under any area of the scale $L$ scales with $R$ as $N_b(L,R) \propto L/R$. Without going into the details of physical mechanism that makes the blocks move, the model assumes that each block randomly "jumps" by $\Delta(R)$, with zero mean and the rms value of the displacement being proportional to $[<\Delta^2(R)>]^{1/2} \propto R^\lambda$ where $\lambda$ is a parameter. Over any given time interval $T$, the number of the jumps $N_j(T,R)$ for various block sizes scales as $N_j(T,R) \propto T/R^\mu$ where $\mu$ is another parameter.

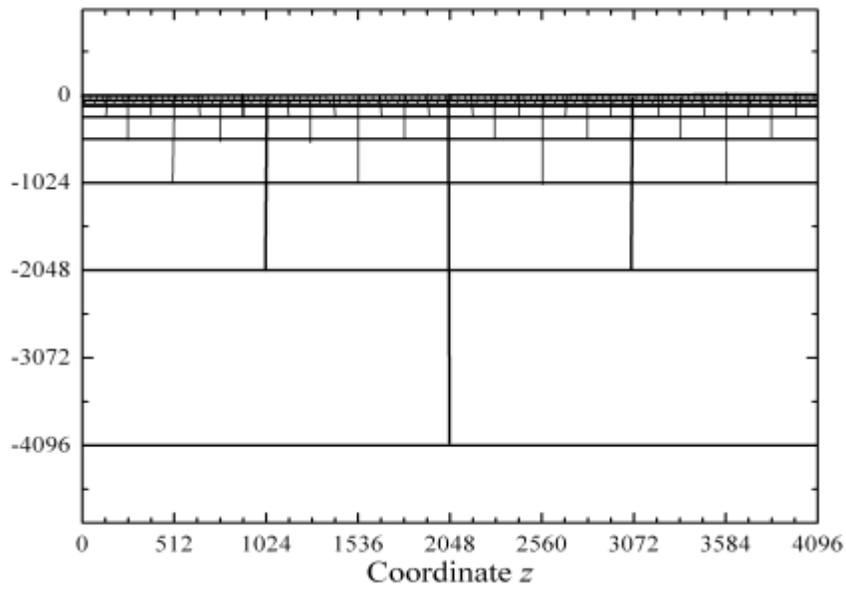

Fig.43: Fractal set of ground blocks (see text)

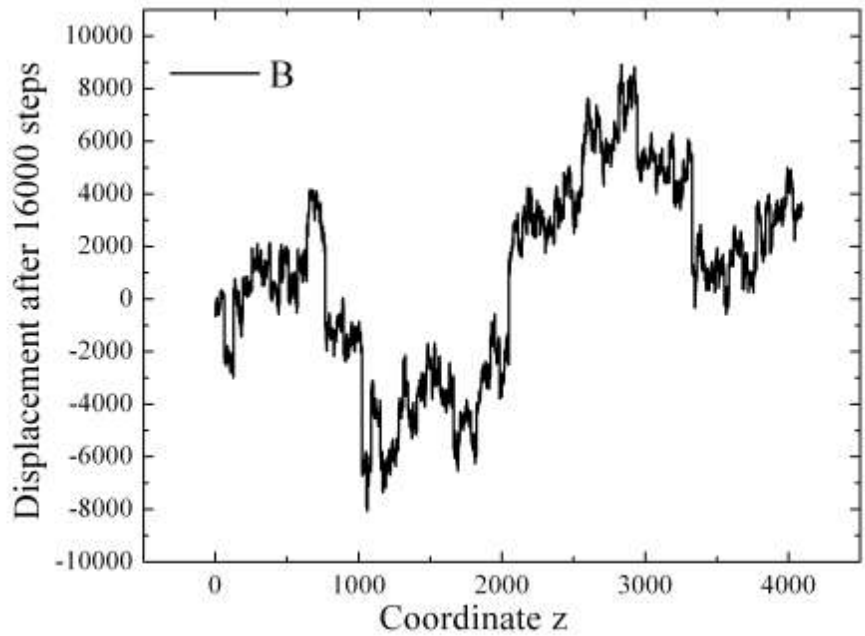

Fig.44: Elevations of the set points after 16,000 steps with parameters of the model $\lambda=\mu=1$.

In computer simulations, each block was considered as a two-dimensional square; the sizes of blocks had been chosen to be $R$ = 1, 2, 4, 8, 16, …2048. The displacement of each the 4096 surface points of the surface is determined as the sum of the displacements of blocks located just *beneath* it. At each time step the blocks having the smallest dimension $R$ = 1 are randomly moved (vertically) with the displacement rms value equal to 1; blocks having $R$ = 2 are displaced after $2^{\mu}$ time steps randomly with rms value of the rms displacement equal to $2^{\lambda}$, etc. Fig.44 shows an example of the resulting profile after 16,000 steps with the parameters of the model $\lambda=\mu=1$.

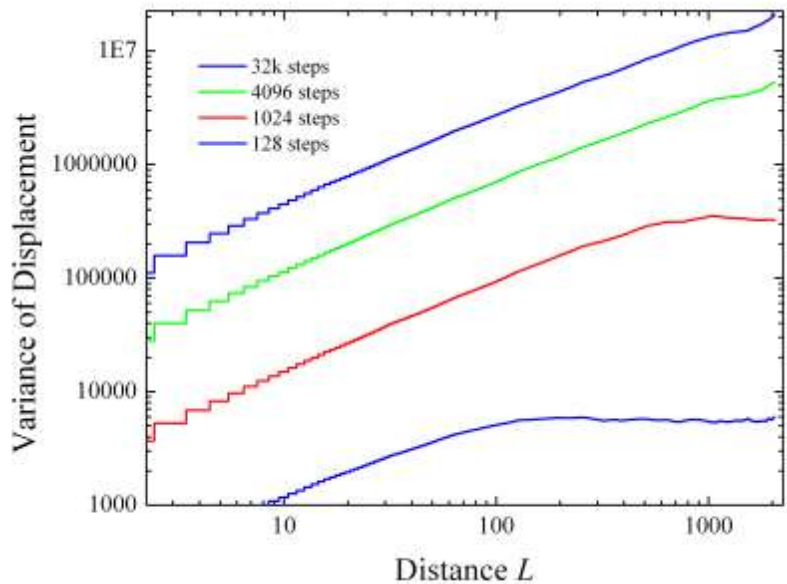

Fig.45: Variance of the displacements vs distance between points for 128, 1024, 4096 and 32,000 steps.

Fig.45 shows the dependence of the variance of the displacement $<dY^2(t,L)>$ on the distance between points for various time intervals $t$=128, 1024, 4096 and 32,000 steps. One can see that after 128 steps, the variance at the distances $L$>128 does not depend on $L$, i.e. $<dY^2(t,L)>\approx const$. The same phenomena occurs after 1024 steps at the distances $L$>1024. Assuming that all the moves are uncorrelated, the average variance of relative position changes for time intervals T can be estimated as:

$$\sigma^2(T,L) = <(Y(t+T,z+L) - Y(t+T,z) - Y(t,z+L) + Y(t,z))^2> =$$

$$= \left( \sum_{blocks}^{R_{max}<L} \sum_{jumps}^{overT} \Delta(R) \right)^2 \propto \sum_{blocks}^{R_{max}<L} R^{2\lambda} \cdot \left( \frac{T}{R^\mu} \right) = \sum_{size\ R=1}^{L} \left( \frac{L}{R} \right) \cdot R^{2\lambda} \cdot \left( \frac{T}{R^\mu} \right) = \quad (11).$$

$$= T \cdot L \cdot \sum_{R=1,2,4,8...}^{L} R^{2\lambda-1-\mu}$$

If the parameter $D$=2$\lambda$-1-$\mu \leq 0$, the sum can be easily calculated and it scales as $\sigma^2(T,L) \propto TL$. If $D$>0, the summation yields $\sigma^2(T,L) \propto TL^{2\lambda-\mu}$. Fig. 46 illustrates how the variance $\sigma^2(T,L)$ scales with $L$ depending on the exponents $\alpha$ and $\beta$. Note, that the time-dependence of the variance can be made different from $\propto T^1$ if the jump frequency scales with time non-linearly $N_j(T,R) \propto T^\nu/R^\mu$. In general, one can conclude that dynamic fractal models like the one we just considered, result in the space-time diffusive motion like one observed in the experimental data discussed in previous sections.

We should note here, that widely accepted Langevin-type stochastic equation for the geological landscape evolution always consider, besides smoothing diffusion and erosion terms, an external stochastic noise source uncorrelated in both space and time and with finite variance – see Ref.[52] for detailed review and discussion. Of course, under these assumptions, the resulted variance scales $\sigma^2(T, L) \propto T$ in the case of no smoothing and no erosion, leaving off any dependence on the distance between the observation points. We believe that such an *ansatz* is basically incorrect as the ground motion noise clearly shows its non-stationary character, certain correlation laws in both space and time and scaling. Besides the ATL-law observations, the fractal statistics of earthquakes [53] repudiates the notion of the stationary uncorrelated noise as the source of the observed ground motion.

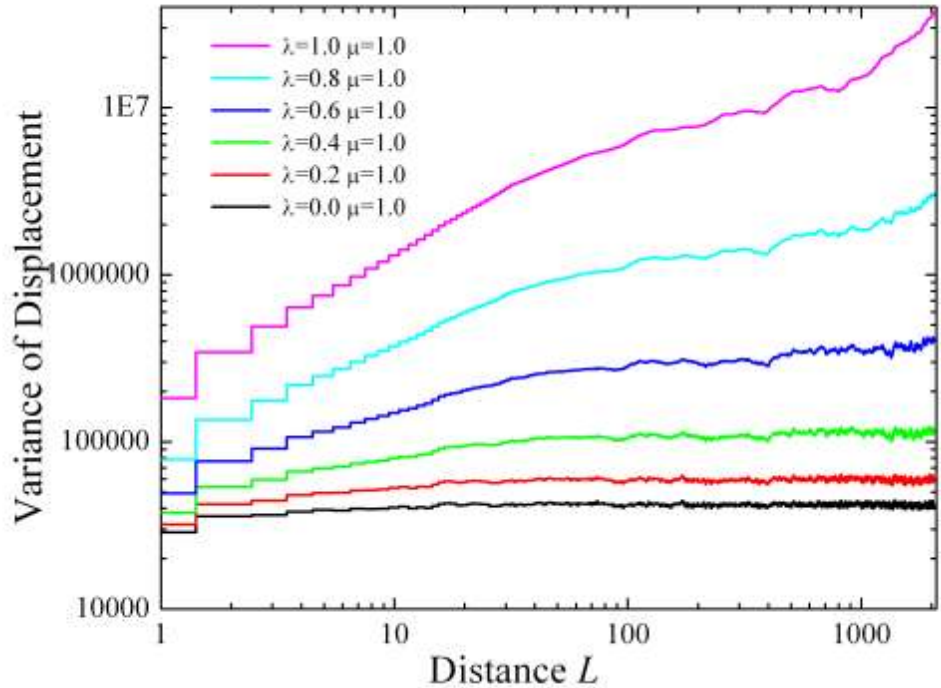

Fig.46: Variance of the displacements vs distance between points after 16,000 steps for different scaling exponents λ and μ (see text).

# 4 Summary

Numerous observations and analysis of the data on slow ground motion presented above reveal the phenomena of simultaneous ground diffusion in space and in time. The diffusion obeys a characteristic fractal law with the ground displacement variance $dY^2$ scaling with time- and spatial intervals $T$ and $L$ as $dY^2 \propto T^\alpha L^\gamma$ with both exponents close to 1 ($\alpha \approx \gamma \approx 1$). The most suitable instruments for studying such a diffusion are arrays of high precision instruments, e.g., Hydrostatic Level Sensors connected by common water pipe and spread over significant area or regular laser tracking of numerous alignment monuments installed in large underground facilities like high energy accelerators. Non-random, systematic movements do often dominate the ground motion but the diffusion components still can be clearly indentified using filtering methods. We believe that present landscape evolution models which assume random stochastic uncorrelated noise as a source of the ground motion are, therefore, incomplete.

Author acknowledges very fruitful long-term collaboration on the development of the HLS probes suitable for the ground motion studies for large accelerators and series of the studies with  them at various places within FNAL-SLAC-BINP team (Batavia-Stanford-Novosibirsk) of B.Baklakov, A.Chupyra, A.Erokhin, J.Lach, A.Medvedko, M.Kondaurov, V.Parkhomchuk, S.Singatulin, A.Seryi, E.Shubin, J.Volk. I am very thankful to the collaborators who provided me with numerous records of raw data for further ground diffusion analysis – Prof.S.Takeda, Drs. N.Yamomote, K.Oide, M.Masuzawa (KEK, Japan), Drs. F.Tecker, M.Mayoud (CERN), the Fermilab's Alignment Group. Over the years I had pleasure to collaborate with Drs. G.Stupakov, A.Seryi, T.Raubenheimer  (SLAC), R,Steining (SSCL), V.Lebedev (FNAL), J.Rossbach (DESY) and C.Montag (BNL) on the theoretical studies of the ground motion effects on high energy particle accelerators.   My special acknowledgements to  Prof. Vasily Parkhomchuk of  Budker INP (Novosibirsk, Russia) who brought my attention to the deep physics issues associated with ground motion at the time when we both were working on the design of a linear *e+e-* collider VLEPP. He also was the first who coined in the term  "ATL-law" while trying to analyze results of long-term measurement of the alignment monuments motion at the site of the UNK collider (Protvino, Russia).  Fermilab is operated by Fermi Research Alliance, LLC under Contract No.  DE-AC02-07CH11359  with the United States Department of Energy.

Rings", in *Proc. of Int. Workshop on Accelerator Alignment IWAA'2004,* Geneva, Switzerland; in [8].